\documentclass[prd,a4paper,eqsecnum,nofootinbib,preprint,floatfix]{revtex4}
\usepackage{epsfig}
\usepackage{colordvi}
\usepackage{graphicx}
\usepackage{bm}
\usepackage{multirow}
\usepackage{enumerate}
\usepackage{amsfonts}
\reversemarginpar
\def\<{\langle}
\def\>{\rangle}
\newcommand{\text}{\rm}
\def\Tr{{\rm Tr}\,}
\def\tr{{\text tr}\,}
\def\Eq#1{Eq.~(\ref{#1})}

\def\tilde{\widetilde}

\def\tr{{\text tr}\,}

\begin{document}

\vspace*{0.3in}
 
\begin{center}
{\large\bf Large-$N$ volume reduction of lattice QCD \\ with adjoint Wilson fermions at weak-coupling.}

\vspace*{1.0in}
\vspace{-1cm}
{Barak Bringoltz\\
\vspace*{.2in}
Department of Physics, University of Washington, Seattle,
WA 98195-1560, USA\\
}
\end{center}

\begin{abstract}
We study the large-$N$ volume reduction of QCD with adjoint quarks regularized on the lattice. Specifically, we use Wilson fermions, and while our $d$-dimensional lattice has $(d-1)$ infinite dimensions, the remaining direction is reduced to a point. We perform a weak-coupling one-loop calculation of the free energy as a function of the holonomy in the reduced direction, and map the regimes in the bare lattice parameter space where the holonomy averages to zero and a $Z_N$-center symmetric configuration is the ground state. For $d=4$ and $N_f=1/2,1$ and $2$ Dirac flavors we see that the center symmetry is intact in a generous regime of the phase diagram that includes the chiral point. Thus we see that large-$N$ volume independence of lattice QCD with adjoint Wilson quarks works at weak coupling. Interestingly, we find that this is true even if the quark mass is quite large, and this opens a path to study the volume reduced large-$N$ pure gauge theory. Finally, we analyze in detail the UV sensitivity of the one-loop potential and show that treating the reduced theory as a $(d-1)$-dimensional effective field theory requires the introduction of certain relevant operators that are a subset of those suggested by \"Unsal and Yaffe to stabilize the center symmetry. This means that different regularizations of the volume-reduced theory can be compared only if one includes these terms in the action.
\end{abstract}


\maketitle

\setcounter{page}{1}
\newpage
\pagestyle{plain}

\section{Introduction}
\label{intro}

The phenomenon of volume independence of $SU(N)$ gauge theories at large-$N$ was first suggested in the influential paper of Eguchi and Kawai \cite{EK}. It is an example of an orbifold equivalence between two gauge theories that are related to each other by projecting out a sub-set of fields. In the Eguchi-Kawai (EK) case, one of the theories (let us denote it by `A') is the gauge theory defined at infinite volume, and the other (theory `B') is the gauge theory with one dimension reduced to a point. Clearly, one can obtain B from A by projecting out from A all the fields that are not invariant to translations in the reduced direction. Theory A can also be obtained from B with a less transparent projection that involves a $Z_N$ global symmetry whose order parameter is the holonomy in the reduced direction (the Polyakov loop). There is a necessary general condition that is required for orbifold projections to become large-$N$ equivalences: the  symmetries defining the projection must be symmetries of the ground states in the corresponding theories  \cite{AEK}. Thus, for volume independence to hold, translation symmetry and $Z_N$ symmetry need to be intact. This was indeed shown already in \cite{EK} (technically this was done using Dyson Schwinger equations of Wilson loops). This can also be seen in the approach of Ref.~\cite{Yaffe_coherent} and, to our knowledge, that reference is also the first to emphasize the role of translation symmetry. 

From the lattice perspective volume independence can have an obvious importance; it provides a potential way to significantly save computational effort in numerically solving the large-$N$ limit of certain gauge theories. Unfortunately, however, and as was discovered in \cite{BHN,MK}, in the physical case of QCD (or for that matter of pure Yang-Mills) the $Z_N$ symmetric vacuum is unstable, and the center symmetry is spontaneously broken in the continuum limit of the volume reduced theory. In retrospect this is not surprising since QCD has a deconfining transitions when its temperature is large enough, or equivalently when its four volume is small enough. Therefore shrinking the volume below some critical value leads to a $Z_N$ deconfining-like transition \cite{KNN}. Several ways to circumvent this problem were suggested over the years and for a review on these we refer to \cite{HN_lat05} and to \cite{DEK}. For an updated state of affairs we refer to the review section of the first reference in \cite{QEK}, and for an intuitive explanation we refer to the second reference of \cite{QEK}. 

From here on we focus on the way Ref.~\cite{AEK} suggested to overcome the EK instability: adding to QCD light adjoint fermions that have periodic boundary conditions in the compactified direction. We shall not discuss at length the relation of this theory to physical QCD, but do wish to mention that it is interesting phenomenologically. Its large-$N$ limit is expected to be equivalent to the large-$N$ limit of QCD with fermions in the antisymmetric representation. The latter theory, in turn, is a generalization of physical, $3$-color,  QCD to a large-$N$ limit in which fermions are truly dynamical and back-react on the gauge fields. Thus, it is a complementary limit to the standard large-$N$ `t Hooft limit, where the fermions are in the fundamental representation and are sub-leading dynamically. For more details we refer to Ref.~\cite{ASV} and to the short summary in the forthcoming publication \cite{BS}.

What the authors of Ref.~\cite{AEK} showed is that in the presence of massless adjoint  quarks, the $Z_N$ symmetry remains intact down to zero volume. This was done by calculating, in weak coupling,  the one-loop potential $V$ as a function of the eigenvalues $\theta_{a=1,2,\dots,N}$ of the holonomy in the small direction. Minimizing $V(\{\theta_a\})$ they saw that a $Z_N$ invariant vacuum, with a uniform eigenvalue density, is the ground state. Note that Ref.~\cite{AEK} was formulated in the continuum, and due to asymptotic freedom, the calculation is reliable because the length of the compactified direction is small compared to the dynamically generated scale in the theory.

A related work to Ref.~\cite{AEK} is that of Ref.~\cite{BBCS}. There, the authors considered a $3D$ theory describing a unitary adjoint scalar field coupled to gauge fields and to adjoint fermions. When projected by a certain center symmetry, the action of this theory was shown to give the action of four dimensional QCD with adjoint fermions defined with one compact discretized direction. All the other spatial directions were kept in the continuum. The one-loop potential of this theory was calculated in perturbation theory, and dimensional regularization with minimal subtraction was used to define certain linearly diverging integrals. The result was surprising and the ground state was found to generically {\em break} the $Z_N$ symmetry --  in contrast to what one might naively expect relying on the analysis of Ref.~\cite{AEK}.

The result obtained in Ref.~\cite{BBCS} prompted us to perform the study we present in this paper and to ask three simple and related questions. First we ask how can one understand the apparent difference between the results of Ref.~\cite{BBCS} and those of Ref.~\cite{AEK}. Second, we ask whether both these studies easily carry to a lattice regularization of the field theory where all directions are discretized. More precisely, we ask if it is actually possible to reduce one of the lattice directions to have a single site, while leaving all other to have an infinite number of sites. This is different from the construction of Ref.~\cite{AEK} in which the compact direction was a continuum (i.e. had an infinite number of sites and a zero lattice spacing). This is also different from Ref.~\cite{BBCS} which had a single site in the compact direction, but where the spatial, three-dimensional, space was  a continuum. 
Third, we wish to know what are the bare lattice parameters for which we can expect such volume reduction to be valid. This is a question of practical importance since if the adjoint quarks need to be very light for reduction to work, then it may be impossible to study them in Monte-Carlo simulations. On the contrary, if the vacuum is $Z_N$ symmetric for heavy fermions as well, then the suggestion of Ref.~\cite{AEK} can be considered as a way to probe pure Yang-Mills at large-$N$.

To answer these questions we generalize the calculation in \cite{AEK} to the lattice. This means we need to pick a fermion discretization and our choice is to work with Wilson fermions. The prime reason is that this is the fermion discretization that we use to study this theory with Monte-Carlo simulations in the companion publication \cite{BS}. Our calculation can also be considered as a generalization of the pioneering Refs.~\cite{BHN,MK} in two ways. Firstly, we generalize them to the case where only one of the euclidean directions is reduced to a point, and secondly, we add dynamical adjoint fermions to the theory. (The first time these works were generalized to a singly reduced direction was already done in \cite{HN}, but since this was not published, we repeat some of its steps here).

The following is the outline of the paper. In Section~\ref{action} we define the action of the theory that we study, and prepare for the one-loop perturbative calculation by fixing axial gauge. In Section~\ref{VG} we calculate the contribution of the gauge fields to the one-loop potential. As mentioned above, this section is very similar to what already appears in \cite{HN}. In Section~\ref{VF} we calculate the contribution of the adjoint fermions to the potential. The form of the resulting potential is summarized in Section~\ref{recap}, where we also remark on several of its properties and on the way its parameters should be chosen in order to get continuum-like equations like the ones that appear in Ref.~\cite{BBCS}. Before we minimize the potential we pause in Section~\ref{EFT} to discuss the way the one-loop potential was calculated in Ref.~\cite{BBCS}, explain the meaning of the results obtained in that work and why they do not signal any problems for large-$N$ volume reduction. This is tied with the way Ref.~\cite{BBCS} treats the UV divergences of the one-loop potential, and we therefore analyze these in detail. In Section~\ref{sym} we return to our lattice regularization and present a map of its phase diagram along the quark mass axis for the case of an isotropic lattice. This is generalized to the case where the lattice spacing in the reduced direction $a_t$ is different from the lattice spacing in the unreduced dimensions $a_s$. In that case we map the phase diagram in the plane of the quark mass and the ratio $a_s/a_t$. We summarize our findings in Section~\ref{summary}.

\section{The action and axial gauge fixing}
\label{action}

The starting point of this section is the action of $d$-dimensional QCD with $N_f$ Dirac fermions regularized on a $d$-dimensional hyper-cubic  lattice. Out of the $d$ space-time dimensions, $d-1$ have an infinite number of sites and one has a single site.  The gauge action is the standard Wilson unimproved action, and the fermions are chosen to be of the Wilson type. Also, we allow for anisotropic lattice spacings and so have two gauge couplings and two hopping parameters. Those that correspond to spatial plaquettes and hopping terms carry the super/subscript ``$s$'' while those corresponding to time-like quantities will have the super/subscript ``$t$''. In principle, one can add an anisotropy in the Wilson coefficients (the $r$), but this is not needed \cite{asym_papers}. The boundary conditions on the compact direction are periodic for all fields. The action is then
\begin{eqnarray}
A &=& A^t_{\rm gauge} + A^s_{\rm gauge}+A_{\rm Fermions}\label{A}\\
A^s_{\rm gauge} &=& \frac{2N}{\lambda_s} \, {\rm Re}\, \sum_{x\atop i<j\in [1,d-1]} \Tr \, U_{x,i}U_{x+i,j}U^\dag_{x+j,i}U^\dag_{x,j} \label{Ags},\\
A^t_{\rm gauge}&=& \frac{2N}{\lambda_t} \, {\rm Re}\, \sum_{x\atop i\in [1,d-1]} \Tr \, U_{x,i}U_{x+i,0}U^\dag_{x,i}U^\dag_{x,0},\label{Agt}\\
A_{\rm Fermions} &=& \bar \psi \,D_W \,\psi,\\ 
D_{xy}&=& \delta_{xy} - \kappa_s \left[\sum_{i} \left( 1 - \gamma_i\right) U^G_{x,i}\, \delta_{y,x+i} + \left(1 + \gamma_i\right)U^{\dag G}_{x,i}\, \delta_{y,x-i}\right]\nonumber \\
&& - \kappa_t \left[\left( 1 - \gamma_0\right) U^G_{x,0}\, \delta_{y,x} + \left(1+\gamma_0\right)U^{\dag G}_{x,0}\, \delta_{y,x}\right]. \label{Dw}
\end{eqnarray}
Here the $(d-1)$-dimensional site index is $x$, and the indices $i,j$ denote the $d-1$ transverse coordinates. We denote the compact direction by $0$. Also, while $U_{x,\mu}$ are the usual link matrices, then $U^G_{x,\mu}$ denote their adjoint representation. Lastly, we denote the `t Hooft couplings $g^2_{t,s}N$ by $\lambda_{t,s}$, and the `hopping' parameters by $\kappa_{s,t}$.

We begin by fixing a gauge in which all the temporal links are diagonal. This is done by writing each temporal link as a unitary conjugation of an $x$-dependent diagonal matrix that we denote by $e^{i\varphi_x} \equiv {\rm diag} \left( e^{i\varphi_1},e^{i\varphi_2},\dots,e^{i\varphi_N}\right)_x$:
\begin{equation}
U_{x,0} = V_{x}\, e^{i\varphi_{x,0}}\, V^\dag_{x}.\label{conjugation}
\end{equation}
If we keep $V_x\in U(N)$ and $\varphi^a \ge \varphi^b$ for $a>b$, then  this is almost a one to one mapping because given a pair of matrices $U_{x,0}$ and $e^{i\varphi_x}$, there is more than one matrix $V_x$ that satisfies \Eq{conjugation}. Specifically, any two matrices $V^{(1)}_x$ and $V^{(2)}_x$ that are related to each other by the right multiplication
\begin{equation}
V^{(1)}_x = V^{(2)}_x \times \Lambda_x,\label{residual}
\end{equation}
with $\Lambda_x$ an arbitrary unitary diagonal matrix, will give, for a given set of eigenvalues $\{\varphi_a\}$, the same matrix $U_{x,0}$. This $\prod_x U(1)^N$ symmetry is a residual gauge symmetry which we will not fix, but do not find it to be important at large-$N$ (see below and \cite{HN}). Using \Eq{conjugation}, the measure of the temporal links on each spatial site becomes
\begin{equation}
\int_{U(N)} DU_{x,0} = \int_{U(N)} DV_x\, \,\left[\prod_{a}\,\int_{-\pi}^{+\pi} \frac{d\varphi_a}{2\pi}\right]\, \prod_{a<b} \sin^2 \left(\frac{\varphi_a-\varphi_b}2\right)_{|_{\varphi^a \ge \varphi^b\atop {\rm for}\, a>b}}\label{measure}
\end{equation}
The next step is to use the $U(N)$ spatial gauge symmetry of the path integral and perform the following change of variables:
\begin{eqnarray}
U_{x,i} &\to& V^\dag_{x} \,U_{x,i} \,V_{x+i},\\
\psi_x &\to& V^G_x\,\psi_x,\\
\bar \psi_x &\to& \bar\psi_x \,V^{\dag G}_x.
\end{eqnarray}
This makes  the path integral and any gauge invariant operators independent of the diagonalizing matrices $V_x$, and we are left with the following gauge-fixed path integral 
\begin{equation}
Z_{\rm gauge-fixed} = \prod_{x,i}\int_{U(N)} DU_{x,i}\, \prod_{a} \int \frac{d\varphi_a}{2\pi} \prod_{a<b} \sin^2 \left(\frac{\varphi_a-\varphi_b}2\right)\, \exp\left[A(U^{ab}_{x,0} = \, \delta_{ab} \, e^{i\varphi^a_x})\right] \label{ZGF}
\end{equation}
This will be our starting point for the one-loop calculation.

\section{The gluonic contribution to the one loop potential}
\label{VG}

In this section we calculate the gluonic contribution to the one loop potential. This is, of course, not the first time it is done, and some related references are \cite{BHN,MK,HN}. 
We begin by noting that the  vacua 
\begin{eqnarray}
U^{ab}_{x,i} &=& \delta^{ab},\\
\varphi^a_{x} &=& \theta^a,
\end{eqnarray} 
are classical maxima of the action that are degenerate for all choices of $\theta$.  In this section we integrate the Gaussian fluctuations around these vacua to break this degeneracy.
We do so by writing 
\begin{eqnarray}
U^{ab}_{x,i} &=& \delta^{ab} + i A^{ab}_{x,i} - \frac12 \left(A^2_{x,i}\right)^{ab} + O(A^3),\\
\varphi^a_{x} &=& \theta^a + \phi^a_x.
\end{eqnarray} 
After some familiar algebra one finds that the second order contribution to the spatial plaquette terms in the action, $A^{s}_{\rm gauge}$, becomes 
\begin{eqnarray}
A^{s,{\rm 2nd \,-\, order}}_{\rm gauge}&=&\frac{2N}{\lambda_s}\,\sum_{x,i<j}{\rm Re}\Tr\, \left[ A_{x,i} A_{x+i,j} + A_{x,i}A_{x,j} + A_{x+j,i} A_{x+i,j} + A_{x+j,i} A_{x,i}  \right.\nonumber \\
&& -\left.A_{x,j}A_{x+j,i}-A_{x+i,j}A_{x,i}-\frac12 A^2_{x,j}-\frac12 A^2_{x+j,i}-\frac12 A^2_{x+i,j}-\frac12 A^2_{x,i}\right],\label{Asgauge}
\end{eqnarray}
and that the temporal part of the action, $A^t_{\rm gauge}$, takes the form 
\begin{eqnarray}
A^{t, {\rm 2nd\,-\, order}}_{\rm gauge}&=&-\frac{N}{\lambda_t}\,\left[ \sum_{a,x,i} \left(\phi^a_{x+i} - \phi_x\right)^2 
 + 4 \sum_{a,b,i} | A^{ab}_{x,i}|^2 \, \sin^2 \left(\frac{\theta^a-\theta^b}2\right)\right]. \nonumber \\\label{Atgauge}
\end{eqnarray}
Let us make the following three remarks:
\begin{itemize}
\item We have not taken into account the way the measure of the spatial link matrices depends on $A$ and so have assumed it to be flat. The leading correction is in fact quadratic in $A$ and will formally contribute to the one loop potential \cite{Rothe}. Nonetheless, because we are at weak coupling, the quadratic pieces that we {\em do} include in Eqs.~(\ref{Asgauge})--(\ref{Atgauge}) are large by factors of $1/\lambda_{t,s}\gg 1$ compared to those quadratic pieces that we neglect. Thus, truncating the action as we do in Eqs.~(\ref{Asgauge}--\ref{Atgauge}) is indeed a consistent approximation.
\item At the quadratic level the $\phi$ fields and the $A$ fields decouple. Because the action of the former does not depend on $\theta$, we ignore it from here on.
\item Fluctuations along the diagonal components of $A$ are completely flat. This reflects the residual gauge symmetry discussed above (see \Eq{residual}). The action of these fields will become non-flat once we appropriately choose a gauge, but because there are only $N$ such fluctuations, integrating over them will make only an $O(N)$ contribution to the effective potential. Since we are mostly focused on the large-$N$ limit of the potential, such a contribution will be $O(1/N)$ suppressed compared to that of the off-diagonal components of the gluons and as a result we can ignore these fields.
\end{itemize}

We are now left with integrating only over the off-diagonal components of the spatial gauge fields $A^{a\neq b}_{x,i}$. To do so we Fourier transform according to 
\begin{equation}
A^{ab}_{x,i} = \int_{|k_i|<\pi} \left(\frac{dk}{2\pi}\right)^{d-1}\, e^{ikx + ik_i/2} A^{ab}_{k,i}, 
\end{equation}
and find the following form of $A^{\rm 2nd\,-\, order}_{\rm gauge}\equiv A^{s, {\rm 2nd\,-\, order}}_{\rm gauge} + A^{t, {\rm 2nd\,-\, order}}_{\rm gauge}$.
\begin{eqnarray}
A^{\rm 2nd \,-\, order}_{\rm gauge} &=& -\frac{N}{\lambda_s} \int \left(\frac{dk}{2\pi}\right)^{d-1} \sum_{ij}  \sum_{a\neq b}\, \left[ A^{ab}_{k,i} \, M^{ab}_{ij}(k) \, A^{ba}_{-k,j}\right],\\
M_{ij}(k) &=& \delta_{ij} \left( \sum_l \sin^2 (k_l/2) - \sin^2(k_i/2) + \frac{\lambda_s}{\lambda_t} \sin^2\left(\frac{\theta^a-\theta^b}{2}\right)\right) \nonumber \\
&&- (1-\delta_{ij}) \, \sin(k_i/2)\sin(k_j/2).
\end{eqnarray}
Since $A^{ab}_{k,i} = \left(A^{ba}_{-k,i}\right)^\star$ there are only $(d-1)\times N(N-1)/2$ independent coordinates for each value of $k$ and we write
\begin{equation}
A^{\rm 2nd-order}_{\rm gauge} = -\frac{2N}{\lambda_s} \int \left(\frac{dk}{2\pi}\right)^{d-1} \sum_{ij}  \sum_{a< b}\, \left[ \left(A^{ab}_{k,i}\right)^\star \, M^{ab}_{ij}(k) \, A^{ab}_{k,j}\right].
\end{equation}
(here we used $M^{ab}_{ij}(k) = M^{ab}_{ij}(-k) = M^{ba}_{ij}(k)$). Integrating over $A$ gives the following contribution to the one-loop potential (the second term comes from the measure of the $\theta$ fields -- see \Eq{measure}):
\begin{equation}
V_{\rm gauge}(\theta) = \sum_{a<b}\,  \int \left(\frac{dk}{2\pi}\right)^{d-1} \, \log \left[\, \det_{ij} \, M^{ab}_{ij}(k)\right] - \sum_{a<b} \log \left( \sin^2 \left(\frac{\theta^a-\theta^b}{2}\right)\right). \label{oneloop_g_1}
\end{equation}

With simple linear algebra one can show that the $(d-1)$-dimensional matrix $\left(M^{ab}(k)\right)_{ij}$ has $(d-2)$ identical eigenvalues equal to $\sum_i \sin^2 (k_i/2) + \frac{\lambda_s}{\lambda_t}\sin^2 \left(\frac{\theta^a-\theta^b}{2}\right)$ and one eigenvalue equal to $\frac{\lambda_s}{\lambda_t}\sin^2 \left(\frac{\theta^a-\theta^b}{2}\right)$ \cite{MK}.\footnote{To see this write $\left(M^{ab}(k)\right)_{ij} = \left(\sum_i \sin^2 (k_i/2) + \sin^2 \left(\frac{\theta^a-\theta^b}{2}\right)\right) \delta_{ij} - \left(\tilde M^{ab}(k)\right)_{ij}$ and diagonalize $\tilde M(k)$. The latter has the structure $\tilde M_{ij} = s_is_j$ with $s_i=\sin(k_i/2)$ and so has one eigenvector proportional to $v_0=(s_1,s_2,\dots,s_{d-1})$, with a corresponding eigenvalue equal to $\sum_i s^2_i$. There are also $(d-2)$ eigenvectors that are orthogonal to $v_0$ and that therefore correspond to zero eigenvalues. This gives the set of eigenvalues for the matrix $M$ discussed in the text.} Dropping $\theta$-independent terms, it is easy to see that the contribution of the latter to \Eq{oneloop_g_1} cancels the second term in \Eq{oneloop_g_1}, and as a result one obtains the following form
\begin{equation}
V_{\rm gauge}(\theta) = \frac{d-2}{2}\,\sum_{a\neq b}\,\int \left(\frac{dp}{2\pi}\right)^{d-1}\log \left\{ \sum_{i} \, \sin^2(k_i/2) + \frac{\lambda_s}{\lambda_t}\sin^2 \left(\frac{\theta^{a}-\theta^b}2\right)\right\}.\label{oneloop_g_2}
\end{equation}
By generalizing the one loop potential in \cite{BHN,MK} this form could have been anticipated in advance. Also, the $L_t\ge 1$ version of \Eq{oneloop_g_2} already appears in \cite{HN}.\footnote{Although with an overall factor of two which seems to be redundant. While unimportant for Ref.~\cite{HN} (that studied only 
the pure gauge case), this factor is crucial for us, as we are comparing the magnitude of \Eq{oneloop_g_2} to the contribution of the fermions to the one-loop potential (see section~\ref{VF}).}
\Eq{oneloop_g_2} is indeed suggestive that, in the $Z_N$ invariant vacuum, $\theta^a-\theta^b$ is what plays the role of the fourth component of the gluon momenta. This is a generic feature of the way single-site reduced models embed the first Brillouin Zone of the large volume theory into color space \cite{BHN,GKPZ}.

\section{The fermionic contribution to the one loop potential}
\label{VF}

To add adjoint Wilson fermions we use the form in \Eq{Dw}.
At one loop we set $U_{x,i} = 1$, and $U^{ab}_{x,0} = \delta_{ab}\,e^{i\theta^a}$. Then $D$ is diagonalized in color space and we have\footnote{Here we work in a basis where the color indices $A$ and $B$ of the adjoint representation link $\left(U^G\right)_{AB}$ are composite indices that correspond to pairs of fundamental indices: $\left(U^G\right)_{AB}\equiv \left(U^G\right)_{ab,cd} = U_{ac} U^\star_{bd}$. This means that the classical configuration of the time-like link is given by
\begin{equation}
\left[\left(U^G\right)_{x,0}\right]_{ab,cd} = \delta_{ac}\, \delta_{bd}  \,e^{i(\theta^a-\theta^b)}\equiv \delta_{ac}\,\delta_{bd}\,e^{i\theta^{ab}}. \nonumber
\end{equation}
.}
\begin{eqnarray}
D_{xy} &=& \delta_{xy} - \kappa_s \left\{\sum_{i}\left[ (1-\gamma_i)\delta_{y,x+i} + (1+\gamma_i)\delta_{y,x-i}\right]\right\}\nonumber \\
&& - \kappa_t\left\{(1-\gamma_0)e^{i(\theta^{a}-\theta^b)} + (1+\gamma_0)e^{-i(\theta^{a}-\theta^{b})} \right\}.\
\end{eqnarray}
In momentum space, $D$ has eigenvalues given by (here we denote $\theta^a-\theta^b$ by $\theta^{ab}$)
\begin{eqnarray}
D_p &=& 1 - \kappa_s \left\{\sum_{i}\left[ (1-\gamma_i)e^{ik_i} + (1+\gamma_i)e^{-ik_i}\right] \right\}\nonumber \\
&& - \kappa_t \left\{(1-\gamma_0)e^{i\theta^{ab}} + (1+\gamma_0)e^{-i\theta^{ab}} \right\},
\end{eqnarray}
which can be written as 
\begin{equation}
D_{p} = 1 - 2\kappa_s \sum_{i} \cos k_i + 2\kappa_t\,\cos\,(\theta^{ab})
 + 2\,i\,\kappa_s \sum_{i} \sin(k_i)\gamma_i +2\,i\,\kappa_t\,\sin(\theta^{ab})\,\gamma_0,
\end{equation}
or as
\begin{equation}
D_{p} = (1 - 2\kappa_s (d-1)-2\kappa_t) + 4\kappa_s \,\left(S + \frac{\kappa_t}{\kappa_s}\sin^2\left(\theta^{ab}/2\right)\right) + 2i\kappa_s \left[\sum_{i}\sin(k_i)\gamma_i + \frac{\kappa_t}{\kappa_s}\,\sin(\theta^{ab})\,\gamma_0 \right].
\end{equation}
Here we defined $S\equiv \sum_{i}\sin^2(k_i/2)$.
In $d$ space-time dimensions, and for $N_f$ Dirac flavors, the determinant of $D_p$ is given by
\begin{eqnarray}
\det D_p &=& \left\{\left((1 - 2\kappa_s (d-1) -2\kappa_t ) + 4\kappa_s \left(S + \frac{\kappa_t}{\kappa_s}\,\sin^2\left(\theta^{ab}/2\right)\right)\right)^2 \right. \nonumber \\
&&\left. + 4\kappa^2_s\left(S_2 + \left(\frac{\kappa_t}{\kappa_s}\right)^2\sin^2\theta^{ab} \right)\right\}^{2^{(d/2-1)}N_f},\label{detDp}
\end{eqnarray}
where we have also defined $S_2\equiv \sum_{i} \sin^2 k_i$. 

Note that the calculation in this section ignores the fermionic zero modes. As shown in Ref.~\cite{0modes} these are important in the supersymmetric case (i.e. when $N_f=1/2$ and for massless quarks). In that case the one loop potential at the vicinity of the $Z_N$ invariant phase is zero, and the subleading, $O(N)$, contribution of the fermionic zero modes is what is left. Since we are not particularly interested in the SUSY theory, and because away from the $Z_N$ invariant state, SUSY is explicitly broken, then we can always neglect these zero modes.

\section{Re-cap and connecting the lattice parameters to the parameters of the continuum calculation of Ref.~\cite{BBCS} }
\label{recap}

To conclude, the following is the 1-loop potential of the full theory:
\begin{eqnarray}
V(\theta) &= &\sum_{a\neq b}\int_{|k_i|<\pi} \, \left(\frac{dk}{2\pi}\right)^{d-1} \, \log\,\left[ \frac{{\cal D}_g(k,\lambda_s,\lambda_t)}{{\cal D}_f(k,N_f,\kappa_s,\kappa_t)}\right],\label{Vtheta}\\ \nonumber \\
{\cal D}_g(k,\lambda_s,\lambda_t)&\equiv&
\left[S + \frac{\lambda_s}{\lambda_t}\sin^2 \left(\frac{\theta^{ab}}2\right)\right]^{(d-2)/2}
,\label{1loop_finalg}\\
{\cal D}_f(k,\kappa_s,\kappa_t)&\equiv& \left\{
\left[(1 - 2\kappa_s (d-1) - 2\kappa_t) + 4\kappa_s \left(S + \frac{\kappa_t}{\kappa_s}\sin^2\left(\frac{\theta^{ab}}2\right)\right)\right]^2 \right.\nonumber \\
&&+\left. 4\kappa^2_s\left(S_2 + \left(\frac{\kappa_t}{\kappa_s}\right)^2\sin^2\theta^{ab} \right)
\right\}^{2^{d/2-1}N_f}
,\label{1loop_finalf}\\
S &\equiv& \sum_{i=1}^{d-1} \sin^2 \left(\frac{k_i}2\right),\\
S_2 &\equiv& \sum_{i=1}^{d-1} \sin^2 \left(k_i\right).
\end{eqnarray}
One way to see that \Eq{Vtheta} makes sense is to set $d=4$ and $N_f=1/2$. In that case the power of the outermost brackets in both \Eq{1loop_finalg} and \Eq{1loop_finalf} is equal to $1$ and so if we take the large-$N$ limit in the $Z_N$ invariant phase, and appropriately tune the bare lattice parameters to have massless quarks in the continuum limit, one can see that $V(\theta)\to 0$, as it should, due to supersymmetry.

\bigskip

Let us now show how to tune the parameters in \Eq{Vtheta} to formally get continuum-like equations like those obtained in Ref.~\cite{BBCS}.\footnote{A related calculation for the pure gauge case was performed in Ref.~\cite{HN}.}  There, the authors work with a model which naively looks like a continuum version of the reduced model that we study here.  To see the connection between our \Eq{Vtheta} and Eq.~(3.14) of Ref.~\cite{BBCS} we re-introduce the spatial and temporal lattice spacings $a_s$ and $a_t$. This is done by writing $k_i = a_s p_i$ and dividing ${\cal D}_g$ and ${\cal D}_f$ by appropriate powers of $a_s$ such that $S$ in the numerator and $S_2$ in the denominator will both get divided by $a_s^2$. Then, if we take $a_s\to 0$ but keep $a_t$ fixed, we have $(4S/a^2_s, S_2/a^2_s) \to \sum_i p^2_i \equiv p^2$. This procedure turns ${\cal D}_g/{\cal D}_f$ into the following form\footnote{Here we also divide the content of the square brackets in the denominator by $4\kappa^2_s$. This gives a potential that differs from the original one by a $\theta$ independent term, which we ignore.}
\begin{eqnarray}
\frac{{\cal D}_g}{{\cal D}_f}&\stackrel{a_s\to 0\atop a_t = {\rm fixed}}{\longrightarrow}&
\frac{\left(p^2 + \frac{4\lambda_s}{\lambda_ta^2_s}\sin^2 \theta^{ab}/2\right)^{(d-2)/2}}{  
\left[
\left(\frac{(1 - 2\kappa_s (d-1) - 2\kappa_t)}{2\kappa_sa_s} + a_sp^2 + \frac{2\kappa_t}{a_s\kappa_s}\sin^2\left(\frac{\theta^{ab}}2\right)\right)^2 + \left(p^2 + \left(\frac{\kappa_t}{a_s\kappa_s}\right)^2\sin^2\theta^{ab} \right)
\right]^{2^{d/2-1}N_f}
}.\nonumber \\ \label{1loop_final_cont}
\end{eqnarray}
Comparing \Eq{1loop_final_cont} with Eq.~(3.14) of \cite{BBCS}, we see that the following relations connect the bare lattice quantities of our calculation (i.e. $\kappa_t,\kappa_s,\lambda_t,\lambda_s$) with the bare quantities of continuum-like calculations of the sort of \cite{BBCS} (i.e. the quark mass in units of the temporal lattice spacing, $a_t m$):\footnote{There may be typos in some of the numerical coefficients of the relevant equations of Ref.~\cite{BBCS}, and for example it seems that the argument of the $(\sin)^2$ term in the equation at the bottom of page $10$ should be divided by $2$, while its coefficient should be multiplied by a factor of $2$. We think this leads to an extra pre-factor of $\sqrt{8}$ to the first term in the square brackets in the equation at the bottom of page $11$ and in Eq~.(3.14). This should not change the conclusions of \cite{BBCS} since this factor strengthens the $Z_N$ destabilizing gluon contribution to the one-loop potential.}
\begin{eqnarray}
a_t &=& a_s \sqrt{\frac{\lambda_t}{\lambda_s}},\\
\kappa_t &=& \kappa_s \frac{a_s}{a_t},\\
\kappa_s &=& \frac{1}{2(d-1) + 2(a_t m +\sqrt{\frac{\lambda_s}{\lambda_t}})}.\label{ks}
\end{eqnarray}

Thus, {\em naively}, one might expect that as we take $\lambda_s/\lambda_t \to 0$ we should recover the results reported in Ref.~\cite{BBCS}. There, the vacuum was seen to be a state that breaks the center symmetry, except for a small, physically uninteresting, window of very heavy quarks with $a_t m \in [0.6,0.8]$.\footnote{We are working with symmetric Wilson parameters that are equal to one.} 

In the next section we discuss the results of Ref.~\cite{BBCS} and show that such an expectation is wrong.

\section{The volume-reduced theory as an effective field theory}
\label{EFT}

In this section we explain why the results of Ref.~\cite{BBCS}  do not signal a problem for large-$N$ reduction of the theory. To show this it is sufficient to focus on the gauge field sector -- later, in Section~\ref{divF}, we will show how fermions modify the discussion. We therefore begin by writing the gluonic contribution of the action studied in Ref.~\cite{BBCS}. It is the naive continuum limit of Eqs.~(\ref{Ags})--(\ref{Agt}) taken only in the spatial dimensions. Thus, it describes a $D$-dimensional gauge theory coupled to a non-linear unitary, adjoint sigma field $\Omega\in SU(N)$ and its action is
\begin{eqnarray}
S^{\rm one-site}(A_\mu,\Omega) &=& \int d^{D}x \,\,\tr\,\left[ \frac1{2g^2}\, \sum_{\mu,\nu=1}^D F^2_{\mu\nu} + \frac{f^2}2 \sum_{\mu=1}^D\left| D_\mu \Omega \right|^2\right],\\\label{AHCGNt1}\nonumber\\
D_\mu \Omega &=& \partial_\mu \Omega  + i\left[A_\mu\, , \Omega \right].
\end{eqnarray}
The case studied in Ref.~\cite{BBCS} had $D=d-1=3$, $g^2=2\lambda_s/(Na_s)$, and $f^2=2N/(\lambda_t a_s)$, but for our discussion it is convenient to leave $g$ and $f$ as free parameters.

Had we allowed for $L_t$ sites in the compact discretized direction, then the continuum gauge fields $A_\mu(x)$ and the unitary field $\Omega$ would acquire an additional index $t=1,2,\dots,L_t$. In that case the field $\Omega_t(x)$ transforms under a bi-fundamental representation of the $SU(N)$ groups generated by $A_\mu(t)$ and $A_\mu(t+1)$. The action is then
\begin{eqnarray}
S^{L_t-{\rm sites}} &=& \sum_{t=1}^{L_t}\,  \int d^{D}x \,\,\tr\,\left[ \frac1{2g^2}\, \sum_{\mu,\nu=1}^D F^2_{\mu\nu}\left(A_\mu(t)\right) + \frac{f^2}2 \sum_{\mu=1}^D\left| D_\mu \Omega(t) \right|^2\right],\label{AHCGNt}\\\nonumber \\
D_\mu\Omega_t &=& \partial_\mu \Omega_t + iA_\mu(t)\, \Omega_t -i \Omega_t A_\mu(t+1).
\end{eqnarray}
An action of the form of \Eq{AHCGNt} has been studied in the context of a low energy effective action for strongly-coupled Higgs models \cite{ABL}, deconfinement of Polyakov loops in four dimensions \cite{BUP}, and of deconstruction of five-dimensional theories as candidates of the standard model \cite{AHCG}. Both  former studies worked with $D=3$ and $L_t=1$, while the latter studied $D=4$ and $L_t\ge 1$.

Below we first discuss several features of $S^{L_t-{\rm sites}}$ that are important in our context (Sections~\ref{nrenorm}-\ref{extra}), then discuss the results of Ref.~\cite{BBCS} (Section~\ref{meaningBBCS}), then analyze the way fermions affect our arguments (Section~\ref{divF}), and finally summarize in Section~\ref{sum}. 

\subsection{Non-renormalizability}
\label{nrenorm}

As noted by Refs.~\cite{AHCG}, the theory defined by \Eq{AHCGNt1} with $D=4$ is non-renormalizable. For $D=3$ this nonlinear sigma model is also not expected to be a renormalizable field theory, at least not in perturbation theory (for example see the remarks in the second reference of \cite{BUP} or in Refs.~\cite{ABL}). 

In fact, dropping the contribution of the gauge fields to \Eq{AHCGNt1} and setting $L_t=1$, we obtain the two-derivative term in the chiral Lagrangian for an $SU(N)\times SU(N)\to SU(N)$ breakdown scheme \cite{GL}. For $L_t>1$ the model is a more complicated sigma model describing the spontaneous breaking of $SU(N)^{2L_t}$ to $SU(N)^{L_t}$ \cite{AHCG}.
While the sigma model for $L_t=1$ and $D=4$ is certainly non-renormalizable (again see \cite{GL}), we are not aware of detailed studies of this issue for the other cases. A simple analysis, however, shows that for the case of interest in this paper ($D=3$ and $L_t=1$) the action in \Eq{AHCGNt1}, when expanded in terms of the `pion' field $\pi$ defined by $\Omega \equiv e^{i\pi}$,
will radiatively generate four derivative terms, which will then generate eight derivative terms, etc., and that all these terms will be linearly UV-divergent. While a similar effect happens at the one-loop level for $D=4$, then the lower UV divergences in  $D=3$ generate the higher derivative terms only at the three-loop level. The one-loop potential in our case, however, is UV sensitive already at one-loop (see below).

In any event, since \Eq{AHCGNt} is non-renormalizable, we can understand it as an effective field theory (EFT).

\subsection{UV sensitivity of the one-loop potential for low values of $L_t$}
\label{UVsensitivity}  

The one-loop potential of our theory contains UV divergences. Denoting a momentum cutoff by $\Lambda$, we should expect a leading $\Lambda^D$ divergence which can be removed by subtracting from the one-loop potential its value at $\theta=0$. Importantly,  however, there are sub-leading divergences that depend on $\theta$, and that  can (and do) affect the breaking scheme of the $Z_N$ symmetry. To show this we first identify the leading divergences in our lattice calculation, where $\Lambda\sim 1/a_s$. 
This is done for $L_t=1$ and $D=3$ in Section~\ref{Nt1D3}. We then discuss in Section~\ref{NtD} the general case of $L_t>1$ and of $D$ spatial dimensions.

\subsubsection{The case of $L_t=1$ and $D=3$}
\label{Nt1D3}

The starting point is to note that the dimensionful one-loop potential is given by multiplying  $V(\theta)$ from \Eq{Vtheta} by $1/(a^{3}a_t)$. Next, we subtract from $V(\theta)$ its value at $\theta=0$ and obtain 
\begin{equation}
\frac1{a_s^{3}a_t}\,\left[V(\theta)-V(0)\right]^{\rm gauge}=\frac{1}{2a_t}\,\sum_{a\neq b} \int_{|p_i|\le \pi/a_s} \left(\frac{dp}{2\pi}\right)^{3} \, \log\left(1 + \frac4{a^2_t}\frac{\sin^2\left(\frac{\theta^{ab}}{2}\right)}{p^2}\right).
\end{equation}
To identify the UV divergence we expand the log in $1/p^2$ and obtain
\begin{equation}
\frac1{a^{3}_sa_t}\left[V(\theta)-V(0)\right]^{\rm gauge}_{\rm UV-divergent} = \frac{2}{a^3_t}\,\sum_{a\neq b}\sin^2\left(\frac{\theta^{ab}}{2}\right)\, \int_{|p|\le \pi/a_s} \left(\frac{dp}{2\pi}\right)^{3}  \frac{1}{p^2}.\label{Vdiv}
\end{equation}
Dropping a $\theta$-independent constant this can be written as
\begin{equation}
\frac1{a^{3}_sa_t}\left[V(\theta)-V(0)\right]^{\rm gauge}_{\rm UV-divergent} \sim  \frac{1}{a^3_t}\,\left| \tr\, \Omega_{\rm classical}\right|^2 \times \frac1{a_s},\label{Vdiv1}
\end{equation}
where we re-identify the classical values of the unitary diagonal holonomy field $\Omega$ as
\begin{equation}
\left(\Omega_{\rm classical}\right)_{ab} = e^{i\theta_a}\, \delta_{ab}.
\end{equation}
Thus we see that the (mass)$^2$ term of the classical holonomy is linearly UV-divergent at one-loop. If we add fermions this will not change, except for one case: for $N_f=1/2$, $m=0$, and in the vicinity of the $Z_N$ invariant vacuum, the one-loop potential is identically zero due to supersymmetry. But away from this vacuum or for other values of $m$ and $N_f$, the $\theta$-dependent linear divergences remain after the introduction of fermions.

\subsubsection{The case of general $L_t$ and $D$}
\label{NtD}

In general, the UV sensitivities of the one-loop potential can be seen from its expansion in terms of bubble diagrams.\footnote{We thank C.~Hoyos for suggesting to us this diagrammatic way of thinking about the one-loop potential.} For that we focus on the quartic interaction between the gauge fields and the $\Omega$ fields -- see the second term in \Eq{AHCGNt1}. For general $L_t$ these interaction terms are given by
\begin{equation}
g^2f^2\, \tr\, \left(\Omega_t A_\mu(t+1) \, \Omega^\dag_t  A_\mu(t)\right) = g^2f^2\,\sum_{abcd} \Omega^{ab}_t \, A^{bc}_\mu(t+1) \, \Omega^{\dag\,cd}_t\,A^{\star\, ad}_\mu(t), \label{vertex}
\end{equation}
which we depict pictorially in Fig.~\ref{vertex}.
\begin{figure}[hbt]
\centerline{
\includegraphics[width=5cm]{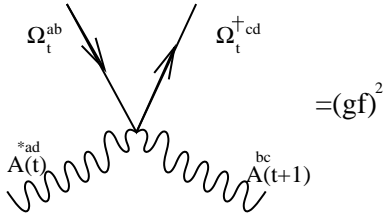}
}
\caption{The vertex of a term of the form of \Eq{vertex}.}
\label{vertex}
\end{figure}

Gauge invariance tells us that the only terms that can be generated radiatively are made out of the Polyakov loop $P(x)$ defined by
\begin{equation}
P(x) = \left(\prod_{t=1}^{L_t} \Omega_t(x)\right),
\end{equation}
and the center symmetry allows only operators of the form
\begin{equation}
\left|\tr\, P(x) \right|^2, \left|\tr\, P^2(x)\right|^2, \left|\tr\,P^3(x) \right|^2,\dots\label{traces}
\end{equation}
(we ignore the fact that operators like $\left(\tr P\right)^N$ are also allowed by the center when the gauge group is $SU(N)$, because at large-$N$ there should be no difference between $SU(N)$ and $U(N)$ and the center of the latter gauge group does not allow such operators).

\begin{figure}[hbt]
\centerline{
\includegraphics[width=5cm]{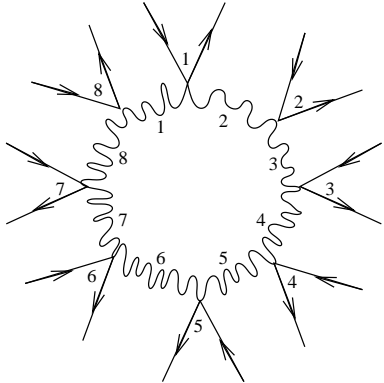}
}
\caption{The bubble diagram that gives rise to the term $|\tr P|^2$ with $L_t=8$. The numbers next to the vertices denote the $t$ index of the $\Omega$ and $\Omega^\dag$ fields on the external legs, while the numbers on the gluon lines denote the $t$ indices of the $A$ fields that flow in the loop.
}
\label{bubbles}
\end{figure}

These symmetry restrictions can be seen by drawing all the possible disconnected bubble diagrams that contribute to $V(\theta)$. For example, the diagram that gives rise to the first operator in \Eq{traces} with $L_t=8$ is given in Fig.~\ref{bubbles}.
A simple counting show that this diagram scales like
\begin{equation}
\frac1{a^D_sa_t}\, \delta V \sim  \frac{\left(fg\right)^{2L_t}}{a_t} \times \int\frac{d^{D}p}{p^{2L_t}},
\end{equation}
which is finite in the $a_s\to 0$ limit so long as 
\begin{equation}
D < 2L_t.
\end{equation}
Ref.~\cite{HN} already showed this by explicitly calculating the one-loop potential for the pure gauge theory as a function of $L_t$.

Thus, for $D=3$ and $L_t=1$ we get the linear divergence discussed above, while for $L_t>1$ one expects no divergences at all. For $D=4$, one expects a quadratic divergence for $L_t=1$, logarithmic for $L_t=2$ and no divergences for $L_t>2$. Indeed, this power counting led Ref.~\cite{AHCG} to consider the $L_t>3$ system as a model for a four dimensional theory that dynamically generates an extra dimension, and whose Higgs mass is not UV sensitive.

Terms in the one-loop potential that correspond to terms in \Eq{traces} with $k\ge 2$ powers of $P$ inside the trace ($k$ windings of Polyakov loops) are generated by bubbles diagrams that are similar to the one in Fig.~\ref{bubbles}, but with $kL_t$ vertices that are ordered cyclically. Thus, the UV contribution to these diagrams is of $O\left(\int \frac{d^Dp}{p^{2kL_t}}\right)$ and is finite as long as 
\begin{equation}
D< 2kL_t.
\end{equation}
 This means that these contributions are all convergent for $L_t=1$ and $D=3$, while for $L_t=1$ and $D=4$, the $k=2$ contribution is logarithmically divergent.

The fact that the $\theta$ dependence of $V(\theta)$ is sensitive to the UV regulator when $L_t$ is small, is, at first glance, quite surprising. Does this contradict the result of Ref.~\cite{AEK} where no divergences were observed? The answer is no and the reason is that the calculation there was defined in the continuum of the compact direction and effectively had $L_t=\infty$. Thus in this case the spatial integrals we discuss above are convergent. Technically this is due to the infinite sum over the Matzubara frequencies corresponding to the compact dimension. Denoting the length of this direction by $R$, this infinite sum yields exponential terms of the form $e^{-R|p|}$ that suppress the higher spatial momenta of $O(1/R)$. 

\subsection{Relevant operators that are missing from the original action in \Eq{AHCGNt}}
\label{extra}

As mentioned above, the  non-renormalizability of the action in \Eq{AHCGNt} tells us that we can see it as an EFT. This means that we can add to the action all possible operators that are allowed by symmetries. The coefficients of these operators are a priori arbitrary and can be chosen at will. As the discussion above shows, we can consider terms of the form
\begin{eqnarray}
\delta S^{(0)}&=& \int d^Dx\left\{B_1 \, \left|\tr P(x) \right|^2 + B_2 \, \left|\tr P^2(x) \right|^2 + \dots\right\},
\end{eqnarray}
with coefficient $B_{1,2,\dots}$ of mass dimension $D$. Other operators we can add have coefficients with negative mass dimensions, for example
\begin{equation}
\delta S^{(2)}= \sum_t \int d^Dx\, \left\{C_1 \left(\tr \left|D_\mu \Omega(t) \right|^2\right)^2 + C_2 \, \left(\tr \left|D_\mu \Omega(t) \right|^2\right)^3 +\dots\right\}.
\end{equation}

\bigskip

While we can choose some of the coefficients of these operators to zero, it is important to understand that at sufficiently high loop order, they may be generated with UV diverging coefficients. In that case, from the EFT point of view, we will need to add them to the action as counter terms whose bare couplings contain UV divergent pieces that cancel the divergences, but also finite pieces that will become new low energy coefficients of the EFT. This is standard in EFT: going to higher loop order means one is sensitive to more low energy constants.

In our case we are working in one-loop, and so need to add the terms to the action that will be generated with UV-divergences at that order. For $D=3$ and $L_t=1$ this means we need to add the following term
\begin{equation}
\delta S_{\rm c.t.} =  \int d^3x\, \left(\frac{c_1}{a_s} + b_1\right)\,\left|\tr P(x)\right|^2.\label{ct}
\end{equation}
The coefficient $c_1$ in \Eq{ct} will cancel the linear divergence of \Eq{Vdiv1}, but the finite low energy coefficient $b_1$ can be chosen at will. Different choices of $b_1$ correspond to different parts of the EFT's space of parameters. Therefore, we see that at one-loop the  EFT in \Eq{AHCGNt1} is defined not only by $g$ and $f$, but also by $b_1$. Clearly, for the purpose of large-$N$ reduction, a preferred choice of $b_1$ will be one that leads to a one-loop potential whose ground state is $Z_N$-invariant. 

For $D$ dimensions and $L_t$ time slices we will need to add terms of the form
\begin{equation}
\delta S_{\rm c.t.} = \int d^Dx\, \sum_{k=1}^{\lfloor\frac{D}{2L_t}\rfloor}\, \left(c_k\Lambda^{D-2L_tk} + b_k\right)\left|\tr P^k(x) \right|^2, \label{cts}
\end{equation}
where $\lfloor x \rfloor$ is the integer part of $x$, and when $D=2L_tk$ there is a logarithmic divergence (for brevity of notations, we assume that each term in \Eq{cts} is radiatively generated only with a leading divergence. This need not be the case in general).

\bigskip

Before we proceed let us emphasize the following two issues:
\begin{enumerate}
\item Terms of the form of \Eq{cts} need to be added to the action {\em only} if one wishes to treat the reduced model as a $D$-dimensional EFT. 

But this is not {\em our} purpose. What we wish to do in this paper is to check whether the theory, as defined in Section~\ref{action}, and for a given set of bare lattice parameters $\lambda_{s,t}$ and $\kappa_{s,t}$, has a vacuum that is $Z_N$ symmetric in weak-coupling. We do not need to think about our theory as an EFT, nor do we need to take its $a_s\to 0$ limit and worry about canceling divergences. If we find that the ground state of our $L_t=1$ reduced model, defined with fixed cutoffs $a_s$ and $a_t$, is $Z_N$ symmetric, then large-$N$ reduction tells us that it is large-$N$ equivalent to the $L_t=\infty$ four dimensional theory defined with the same field content, the same regularization, and the same cutoffs. To remove the four-dimensional cutoffs we will then tune the lattice parameters according to their four dimensional RG flow. According to the large-$N$ equivalence paradigm, this should be done only after taking the large-$N$ limit. Thus in this approach (which, in fact, is the standard one used to justify large-$N$ reduction non-perturbatively \cite{EK,BHN,GK}) we do not add any counter terms to the action.\footnote{There is of course another reason to add terms of the form $|\tr P^k|^2$ with $k\ge 1$; such terms can get rid of unwanted center symmetry breakdown in regimes of the lattice parameter space where it surely happens (see details in Ref.~\cite{DEK} and below). In that case, however, they are not considered as counter terms: their coefficients need not be tuned like $O(1/a_s)$ and the choice of the power $k$ is not dictated by the considerations in Section~\ref{UVsensitivity}.}

In contrast, in Ref.~\cite{BBCS}, the authors do treat the reduced  theory as an EFT. Specifically, minimally subtracted dimensional regularization was chosen to get rid of the UV divergences. For the motivation of treating the theory in this way see the introduction of that paper.

\item It is interesting that the terms in \Eq{cts}, that we need to add to the action from the point of view of EFT, are a subset of the terms that \"Unsal and Yaffe suggested to add to the EK model in order to stabilize its $Z_N$ invariant vacuum \cite{DEK}. As shown there, when projected back to $4D$, these terms change the gauge theory in a way which is only sub-leading at large-$N$. 
\end{enumerate}

\subsection{Connecting the results obtained in Ref.~\cite{BBCS} to other regularization schemes}
\label{meaningBBCS}

The lesson of the previous sections is that if we view the reduced theory as a $D$-dimensional EFT, then  the action describing it is {\em not} given by \Eq{AHCGNt1}; the latter has certain terms missing. In particular, for $D=3$ and $L_t=1$, the most general EFT is given by 
\begin{equation}
S^{\rm 3D} = \int d^{3}x \,\left\{\,\tr\,\left[ \frac1{2g^2}\, \sum_{\mu,\nu=1}^D F^2_{\mu\nu} + \frac{f^2}2 \sum_{\mu=1}^D\left| D_\mu \Omega \right|^2  \right] + \left(c_1\Lambda + b_1 \right) \left| \tr P(x)\right|^2\right\}.\label{S3D}
\end{equation} 
 Here we need to tune $c_1$ in a regulator-dependent way, and to choose $b_1$ as we wish -- different choices correspond to different low energy constants of the EFT. 

Since different regularization schemes subtract the UV divergences in different ways, then one can compare regularization schemes only if one adds \Eq{ct} to the action. For example, the regulator used in Ref.~\cite{BBCS} was minimally subtracted dimensional regularization (MSDR). This regulator sets power law divergences, like the linear divergences in the one-loop potential, to zero, and essentially replaces the UV divergence of $V(\theta)$ by finite $\theta$-dependent functions. This fact makes MSDR a `dangerous' regularization scheme in our context -- it automatically subtracts the infinity, and in contrast to other regularization schemes, it does not make the necessity of the counter terms in \Eq{S3D} manifest. In fact, this was already noted in Ref.~\cite{Luscher}.\footnote{See Appendix $C$ there. We thank M.~\"Unsal for bringing this reference to our attention.} 

Indeed, because MSDR sets the linear infinity to zero, then there is nothing to subtract and the infinite piece of the counter term ($c_1$ in \Eq{S3D}) is fixed to zero in this regularization. Since, however, the last term of \Eq{S3D} was not considered in Ref.~\cite{BBCS}, then $b_1$ was implicitly set to be zero as well. Consequently, the resulting $V(\theta)$ was minimized in a subspace of the EFT full parameter space, and this subspace is not special in any sense (the point $b_1=0$ is not protected by any symmetry). Therefore the fact that $V(\theta)$ was found to break the center symmetry in a physically relevant region of the EFT parameter space is a result that is particular for the choice $b_1=0$ and  may certainly change once one explores other choices for $b_1$ which correspond to other choices of regulators.\footnote{The action Ref.~\cite{BBCS} studied also contained fermions and, compared to \Eq{S3D}, it has more terms in its action (which require an one additional counter term). In the next subsection we discuss this issue, but for the arguments in the current subsection it is not essential.}

For example, we can choose any $3D$ lattice regulator for \Eq{S3D} (such as the standard Wilson action, or an `improved' one), or variants of dimensional regularization like power divergence scheme \cite{PDS}. If one does not add the counter terms to the actions of these regulators, then each regulator makes its own implicit choice for $b_1$. As an example, let us choose the regulator to be a lattice and use the standard Wilson action. This is in fact the same action that we used in Section~\ref{action}. (While in Section~\ref{extra} we emphasized that our lattice reduced model is not taken to be a regulated EFT, we can momentarily depart from this point of view, and use it as one). 
What are the values of $c_1$ and $b_1$ that our lattice calculation chooses when viewed as an EFT? In general, if we denote by $c'_1$ the coefficient of the  $1/a_s$ term that multiplies $|\tr P|^2$ in the lattice result for $V(\theta)$, then we need to set $c_1$ from \Eq{ct} to be $-c'_1$. Next, since the action that we use in our lattice calculation  is only Eqs.~(\ref{Ags})--(\ref{Agt}), then had we treated it as an EFT means that we actually chose $b_1=-c_1/a_s=+c'_1/a_s$.

Our message in this section is that the absence of counter terms leads the regulator to implicitly choose different values of the low energy constant $b_1$. This choice is determined by the details of the regulator and so even if we fix  the physical parameters $f$ and $g$, then there is  no reason why the two regularizations will yield the same physical result. In particular, it is quite possible that while one regularization sees a $Z_N$ invariant vacuum, then the other concludes that the $Z_N$ symmetry is intact, even if $f$ and $g$ are the same in both. To get identical physical results in any two regularizations, one will need to explcitly add the counter term to their action, and tune the values of $b_1$ in both regulators in an appropriate manner.

\subsection{Effect of adjoint fermions on the UV sensitivity of the one-lop potential}
\label{divF}

In this section we show how the presence of the adjoint fermions modifies the discussion above. The modifications are two-fold. First, because of the fermion propagators are of $O(1/p)$, the one-loop divergences can be  of a higher degree. Second, because the Dirac operator of the fermions is a first derivative in the compact direction, which becomes $\sim \sin(\theta^{ab})$ in the reduced model, then the theory will radiatively generate operators of the form $|\tr P^2|^2$ as well as $|\tr P|^2$. 

These facts can be seen in two ways. From arguments of the sort of Section~\ref{Nt1D3} we see that the UV sensitive $\theta$-dependent pieces of the fermionic contribution to the one-loop potentials are given by 
\begin{equation}
\frac1{a^3_sa_t}\left|V(\theta)-V(0)\right|^{\rm fermions}_{\rm UV-divergent} \sim \sum_{ab}\left[ \sin^4\left(\frac{\theta^{ab}}2\right),\,\, {\rm or}\,\, \sin^2\left(\frac{\theta^{ab}}2\right), \,\, {\rm or}\,\, \sin^2\left(\theta^{ab}\right)\right]\times \int \,\frac{d^3p}{p^2}, 
\end{equation}
and a simple rearrangement of the color indices tells us that both $|\tr P^2|^2$ and  $|\tr P|^2$ are generated with a linearly diverging coefficient. This means that the EFT needs to also contain the $|\tr P^2|^2$ operators, and  is thus defined not only by the quark mass and by $f$, $g$ and $b_1$, but also by the values of a new  low energy constant $b_2$ that comes from a counter term of the form
\begin{equation}
\delta S^{\rm fermions}_{\rm c.t.}=\int d^3x \, \left(\frac{c_2}{a_s}+b_2 \right)\, |\tr P^2(x)|^2.
\end{equation}

Arguments relying on the structure of bubble diagrams, like those of Section~\ref{NtD}, can also be used. The bubble diagrams are now generated by replacing the vertex in Fig.~\ref{vertex} with a corresponding vertex that connect two quarks and two $\Omega_t$ matrices, and by replacing the gluonic loop of Fig.~\ref{bubbles} by a quark loop. Since the trace over the Dirac gamma matrices will null all diagrams with an odd number of vertices we find that for general $D$ and $L_t$ one needs to generalize \Eq{cts} to a sum over $k=2,4,6,\dots,k_{\rm max}$ where $k_{\rm max}=\lfloor \frac{D}{L_t} \rfloor$ if $\lfloor \frac{D}{L_t} \rfloor$ is even and  $k_{\rm max}=(\lfloor \frac{D}{L_t} \rfloor-1)$ if $\lfloor \frac{D}{L_t} \rfloor$ is odd.

\subsection{Summary}
\label{sum}

A simple and important result of the previous subsections is the following. If one wishes to treat the reduced model as a three-dimensional EFT, then  $V(\theta)$ calculated for the action in \Eq{A} (if we regularize the EFT with a $3D$ lattice) or for the action \Eq{AHCGNt1} generalized to include fermions (if we choose MSDR), is missing the following terms
\begin{equation}
\delta V_{\rm missing} \sim  b_1 |\tr P |^2 + b_2 |\tr P^2|^2. \label{missing}
\end{equation}
Different values of $b_1$ and $b_2$ correspond to different points in the parameter space of the EFT. Thus, to show that for given values of physical parameters like $a_tm$ and $N_f$, the EFT spontaneously breaks the $Z_N$ center symmetry, we need to verify that there is no combination of $b_1$ and $b_2$ that can make the ground state of $V(\theta)$ center invariant. Since this procedure was not included in the analysis of Ref.~\cite{BBCS}, and instead, the regulator used there implicitly fixed $b_1=b_2=0$, then it is certainly possible that for the same values of $m$ and $N_f$ there is a different point in the plane spanned by $b_1$ and $b_2$ for which $V(\theta)$ has a $Z_N$ invariant ground state. 

Let us show that this is very plausible. First, note that because different regulators subtract the UV divergences in a way that differs by finite pieces, then two regularization schemes will give the same physical results for different values of $b_{1,2}$. The differences between the regulator-dependent values is, however, finite. Next, a straight-forward generalization of the discussion in Section~\ref{meaningBBCS} tells us that in the absence of $\delta V_{\rm missing}$, the lattice regulator effectively fixes $b_{1,2}= c'_{1,2}/a_s$ (here $c'_{1,2}$ are the coefficients of the terms that multiply $|\tr P|^2$ and $|\tr P^2|^2$ in $V(\theta)$ and that scale like $1/a_s$ at small $a_s$). Finally, in Section~\ref{asym} we show that for small $a_s$, the $Z_N$ symmetry is generically {\em unbroken} in our lattice calculation. This means that $c'_{1,2}>0$ (otherwise there would be an instability).  Therefore, if instead of letting MSDR fix $b_{1,2}=0$ for us, we fix these coefficient such that the $|\tr P|^2$ and $|\tr P^2|^2$ terms in $V(\theta)$ have the same coefficients as they do on the lattice, we will find that the $Z_N$ symmetry is intact in MSDR as well. In particular, for small $a_s$ this means fixing $b_{1,2}$ to have  large and positive values. This of course is not surprising: Ref.~\cite{BBCS} report a $Z_N\to Z_2$ symmetry breaking at $m=0$ and $N_f=1$, but by increasing $b_1$ and $b_2$ to large positive values this surely will change and a $Z_N$ symmetric ground state will probably emerge.

\bigskip

In the next section we depart from the EFT point of view, and simply study the lattice one loop potential as a function of its bare parameter space. Thus, we set the counter terms to zero and so fix $b_{1,2}=0$. As a prelude to the full study of the phase diagram we first analyze the case of $a_s=a_t$ in Section~\ref{sym}, and indeed find that the $Z_N$ symmetry seems to be intact in the chiral limit. Next, in Section~\ref{asym}, we fully explore the phase diagram of the potential \Eq{Vtheta} and find that the regime where the $Z_N$ symmetry is intact becomes extended when $a_s/a_t$ is allowed to be different from one. Since in both cases, we find that the $Z_N$ is intact in the physically relevant regimes, we do not continue to ask what happens when we make  $\delta S_{\rm c.t.}$ nonzero.

\section{The phase structure along the $am$ axis: the case of symmetric lattice spacings}
\label{sym}

For symmetric lattice spacings we set $\lambda_t=\lambda_s$ and $\kappa_s=\kappa_t\equiv \kappa$ into \Eq{Vtheta}, and turn to compare the values of the one loop potential for three vacua that realize the $Z_N$ symmetry differently. We do so for the physically interesting case of $d=4$ (although below we present the analytic formulas for general $d$) and for different values of $\kappa$ (recall that at tree level massless fermions are  obtained for $\kappa=1/2d$ -- see \Eq{ks}). The vacua we considered are:
\begin{itemize}
\item A vacuum denoted by $\O$ that completely breaks the $Z_N$ symmetry. Here we set
\begin{equation}
\theta^{ab}=0,
\end{equation}
and find
\begin{equation}
V_{\O}/N^2 = \int \, \left(\frac{dp}{2\pi}\right)^{d-1} \, \log \left\{ \frac{S^{(d-2)/2}}{  
\left[
\left((1 - 2\kappa d) + 4\kappa S\right)^2 + 4\kappa^2S_2
\right]^{2N_f}
}\right\}. \label{V0}
\end{equation}

\item  A vacuum that preserves the $Z_N$ symmetry. Here we set
\begin{equation}
\theta^{ab} = \frac{2\pi(a-b)}{N}.
\end{equation}
Substituting this into the one-loop potential and using
\begin{equation}
\frac{1}{N^2}\sum_{a\neq b} \, f\left(\theta^{ab}\right) \stackrel{N\to\infty}{\longrightarrow}  \int_{-\pi}^\pi \frac{dk_0}{2\pi}\, f(k_0)\label{embed}
\end{equation}
and find
\begin{equation}
V_{Z_N}/N^2 = \int \, \left(\frac{dp}{2\pi}\right)^d \, \log \left\{ \frac{\tilde S^{(d-2)/2}}{  
\left[
\left((1 - 2\kappa d) + 4\kappa \tilde S\right)^2 + 4\kappa^2\tilde S_2
\right]^{2N_f}
}\right\}.\label{VN}
\end{equation}
Note that here the integration is over a $d$-dimensional Brillouin Zone and, correspondingly, $\tilde S$ and $\tilde S_2$ are defined as sums over $d$ terms:
\begin{equation}
\tilde S = \sum_{\mu=1}^d \sin^2k_\mu/2 ,\,\, {\rm and}\,\,\tilde S_2 = \sum_{\mu=1}^d \sin^2k_\mu.
\end{equation}

Indeed, \Eq{embed} is the way large-$N$ reduction embeds space-time into color space, and decompactifies the reduced direction. It is easy to check that \Eq{VN} is exactly the one-loop potential one would obtain if one had an infinite lattice theory in all directions. 

\item We also studied a ground state with a $Z_2$ symmetry, i.e. that has
\begin{equation}
\theta^{a}=\left[
\begin{array}{cl}
0 & \quad a\in [1,N/2],\\
\pi & \quad a\in [N/2+1,N].
\end{array}
\right.
\end{equation}
This means that out of the $N^2$ pairs of indices $a$ and $b$ there are $N^2/2$ that have a potential equal to $V_{\O}/N^2$, and the rest have an interaction given by
\begin{equation}
\Delta V_{Z_2} = \int \, \left(\frac{dp}{2\pi}\right)^{d-1} \, \log \left\{ \frac{ (S+1)^{(d-2)/2}}{  
\left[
\left((1 - 2\kappa d) + 4\kappa  (S+1)\right)^2 + 4\kappa^2 S_2
\right]^{2N_f}
}\right\},\label{V2}
\end{equation}
since for these $\theta^{ab}=\pi$. Thus we see that this ground state has an energy of 
$V_{Z_2}/N^2 = \frac12 \left( V_{\O} + \Delta V_{Z_2} \right)/N^2$.

\end{itemize}
An obvious uncertainty in our calculations is that we have only compared energies of the three vacua described above, and there may be other relevant vacua that we are ignoring. 

\bigskip

To obtain the phase diagram along the $\kappa$ axis, we scanned the values of $V_{Z_N}, V_{Z_2}$, and $V_{\O}$, in $\kappa \in [0,2]$ and for $N_f=0.5,1,2$.\footnote{The case of $N_f=1/2$ corresponds to a single Majorana fermion whose one-loop potential should vanish in the continuum limit of the chiral theory. The reason is simple: in that limit the theory is supersymmetric and the bosonic perturbative contribution must be canceled by the fermionic one. As is well known \cite{AEK}, in the absence of the one-loop potential, non-perturbative instanton effects become important and make the ground state $Z_N$-symmetric. On the lattice, however, and away from the chiral limit, super-symmetry is broken, and we expect the one-loop potential to determine the ground state at sufficiently weak couplings. In that case the instanton effects should be exponentially small.} The integrations over $k$ were done numerically with a trapezoid method whose grid had a resolution of $2\pi/L$ in each direction with $L=50,100,150$. To obtain the $L\to \infty$ limit of these numerical integrations we performed linear extrapolations of $V_{Z_2}$, $V_{\O}$ and $V_{Z_N}$ in $(2\pi/L)^p$ with $p$ equal to the dimension of the Brillouin zone appearing in Eqs.~(\ref{V0}),~(\ref{V2}), and~(\ref{VN}) ($p=3$ in the former two and $p=4$ in the latter). 

\begin{figure}[hbt]
\centerline{
\includegraphics[height=6cm,width=10cm]{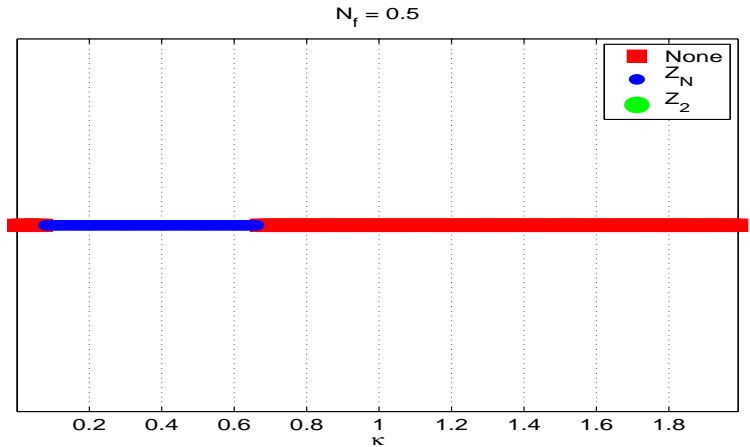}
}
\caption{A map of the phase diagram symmetric lattice spacing and $N_f=1/2$ (a single Majorana fermion in the continuum of the $Z_N$ invariant phase) as a function of $\kappa$ ($\kappa=1/8$ is the chiral point and it has $Z_N$ symmetry intact).}
\label{sym05}
\end{figure}

\begin{figure}[htb]
\centerline{
\includegraphics[height=6cm,width=10cm]{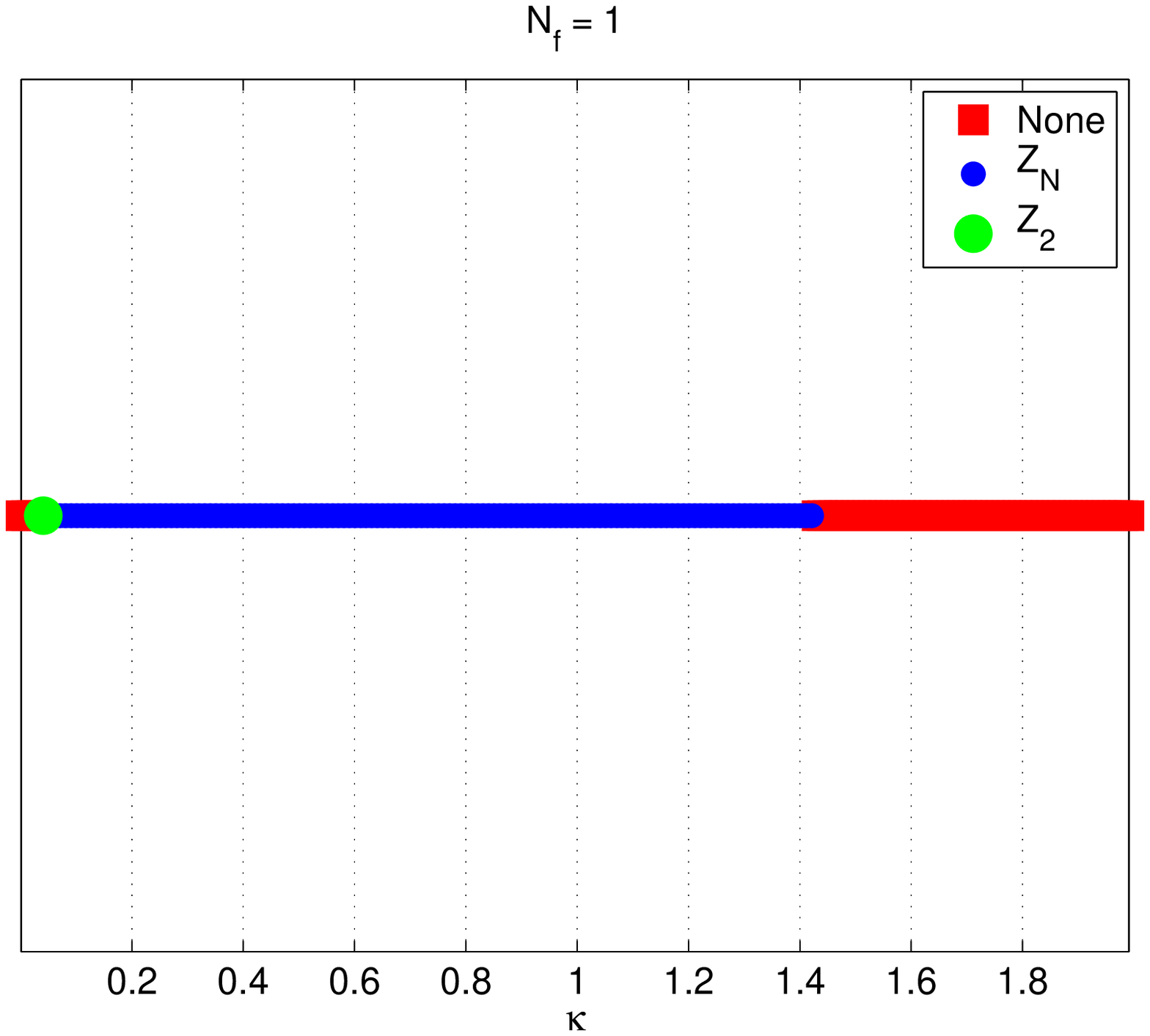}
}
\caption{As in Fig.~\ref{sym05} but for $N_f=1$.}
\label{sym1}
\end{figure}

\begin{figure}[hbt]
\centerline{
\includegraphics[height=6cm,width=11cm]{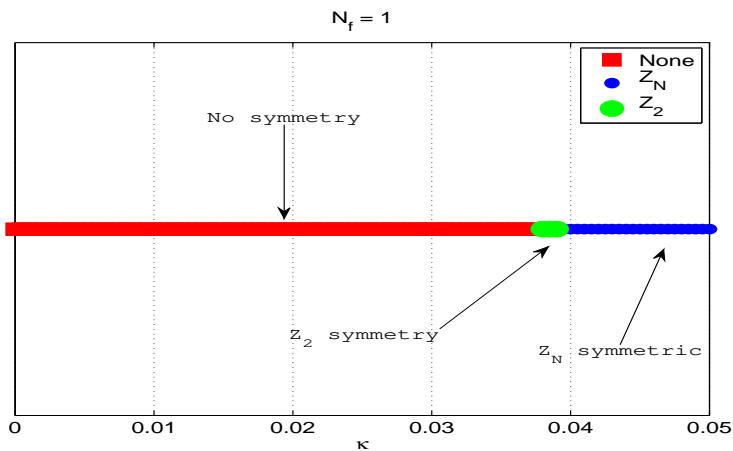}
}
\caption{As in Fig.~\ref{sym1} but zooming on the regime of $\kappa \in [0,0.05]$.}
\label{sym1_1}
\end{figure}

\begin{figure}[hbt]
\centerline{
\includegraphics[height=6cm,width=10cm]{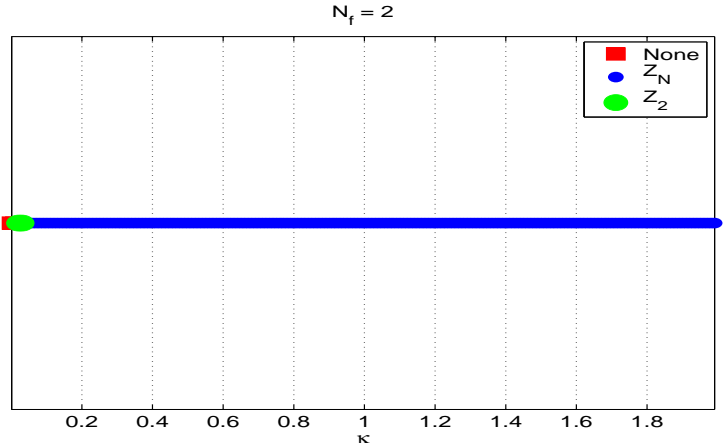}
}
\caption{As in Fig.~\ref{sym05} but for $N_f=2$.}
\label{sym2}
\end{figure}

In  Figs.~(\ref{sym05}),(\ref{sym1}), and (\ref{sym2}) we present maps of the phase space along the $\kappa$ axis for the cases $N_f=0.5,1,2$. For $N_f=1$ we also zoom, in Fig.~\ref{sym1_1}, on the regime of small $\kappa$ (that particular data set was generated for a single value of $L=80$, but, in general, the variation with $L$ was seen to be weak as long as we restrict to $a_s/a_t=1$, as we do in this section).

As the figures show, close to $\kappa=0$ one find $Z_N$ symmetry breakdown, which makes way to a $Z_N$ invariant state when we increase $\kappa$. Surprisingly, this happens at quite small values of $\kappa$: at $\kappa\simeq 0.06$ for $N_f=0.5$, at $\kappa\simeq 0.04$ for $N_f=1$, and at $\kappa\simeq 0.01-0.04$ at $N_f=2$. In terms of the bare quark mass in lattice units these values correspond to 
$am\simeq 50-4$. At even larger values of $\kappa$, the $Z_N$ symmetry breaks again. This, however, is less important since that regime corresponds to the so called `super-critical' regime of Wilson fermions which one needs to avoid in lattice simulations (it is not in the same universality class of QCD \cite{Aoki}).

\bigskip

Importantly, the chiral point (and a generous vicinity thereof) is at $\kappa=1/8$ and this is within the $Z_N$ symmetric phase for all choices of $N_f$. Also, as anticipated, the $Z_N$ symmetric phase becomes more extended with increasing number of flavors.

\section{The phase diagram in the plane of $a_tm$ and $a_s/a_t$.}
\label{asym}

We now turn to map the phase diagram in the plane of $a_tm$ and $a_s/s_t$. As in Section~\ref{sym}, we performed the numerical integrations with a fixed grid in momentum space (which in this section was set to be $2\pi/L$ with $L=60,90$) and extrapolated to zero grid spacing (see discussion in previous section). In contrast to the case of $a_s/a_t=1$ discussed above, when $a_s/a_t$ is small this extrapolation is important to perform and sticking to a fixed value  of $L$ can result in an erroneous phase diagram. 
\begin{figure}[hbt]
\centerline{
\includegraphics[width=10cm]{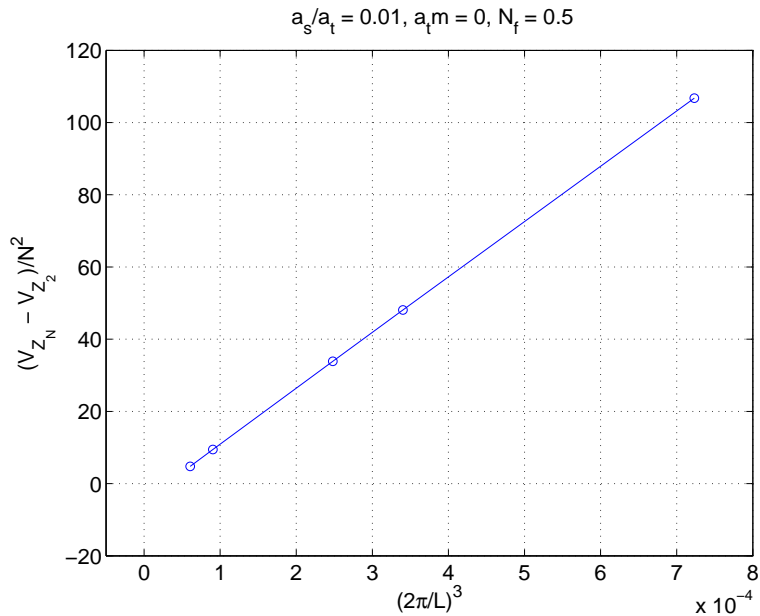}
}
\caption{The difference in energy between the $Z_N$ invariant state and the $Z_2$ invariant state for $N_f=1/2$, $a_tm=0$ and $a_s/a_t=0.01$ versus the momentum space resolution used to perform the numerical integrations over the Brillouin zone. At $L\to \infty$ the difference is negative and the $Z_N$-invariant vacuum is energetically preferable.
}
\label{V_extrapolate05}
\end{figure}
\begin{figure}[hbt]
\centerline{
\includegraphics[width=10cm]{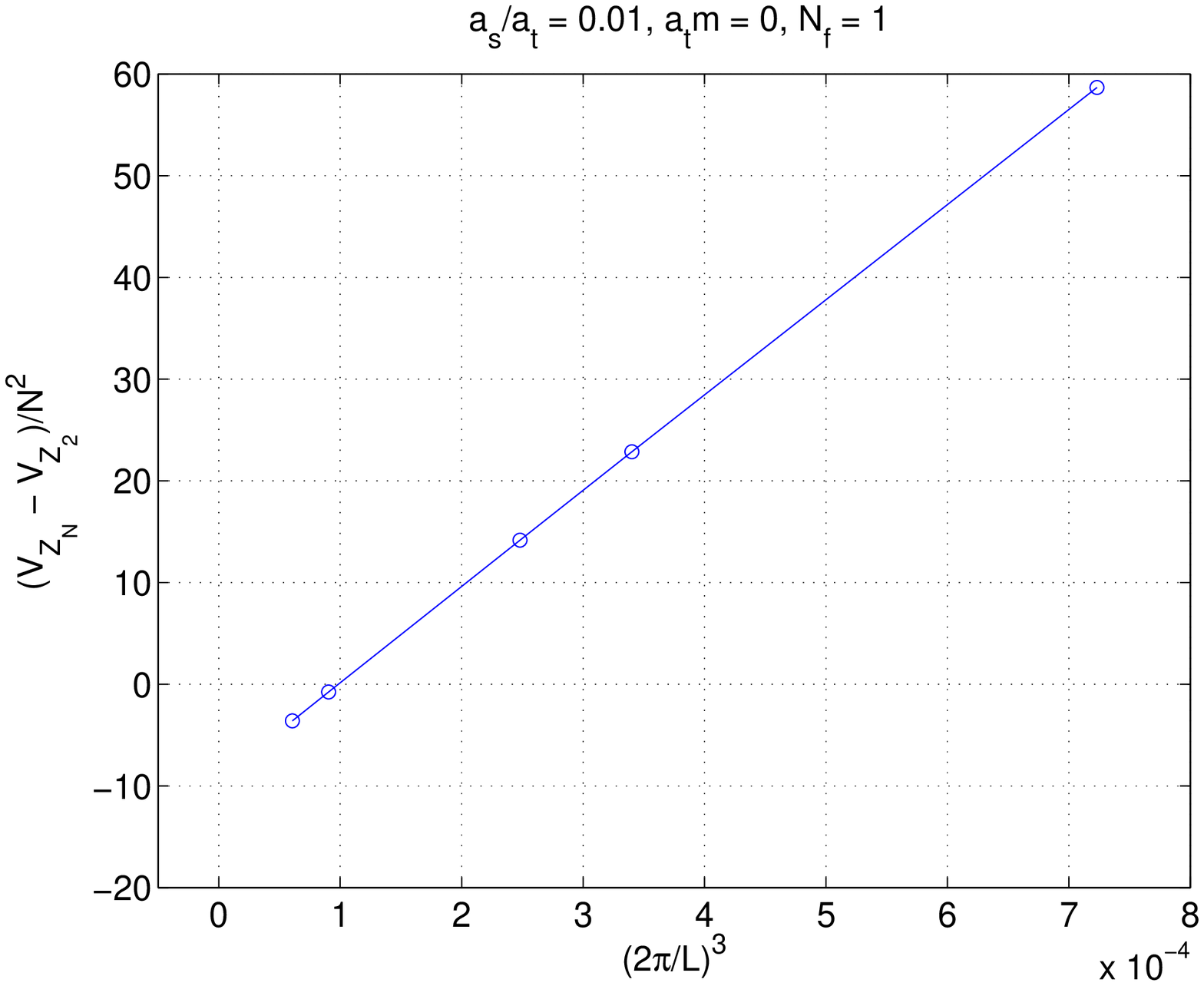}
}
\caption{The same as in Fig.~\ref{V_extrapolate05}, but for $N_f=1$. 
}
\label{V_extrapolate1}
\end{figure}
\begin{figure}[hbt]
\centerline{
\includegraphics[width=10cm]{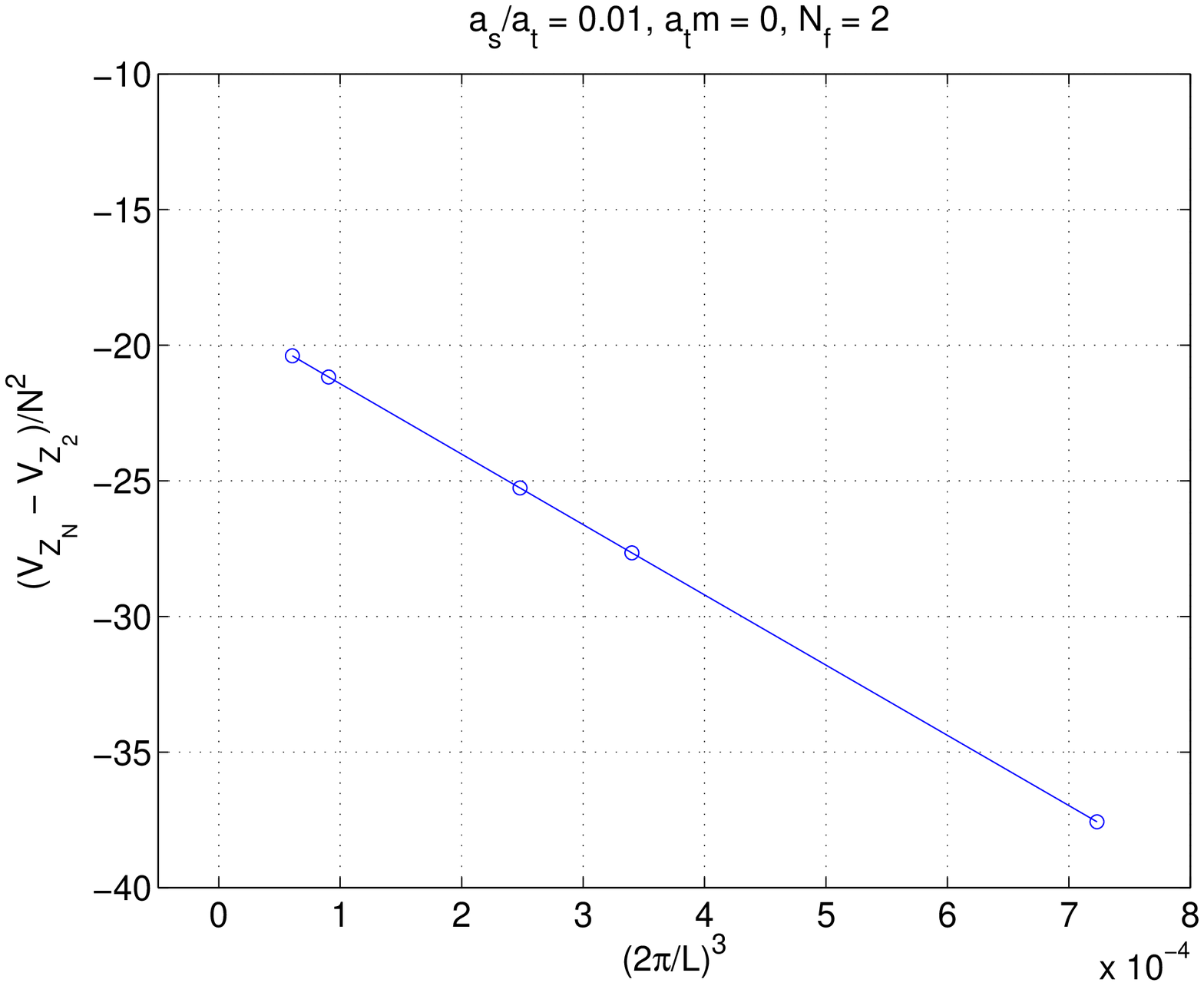}
}
\caption{The same as in Fig.~\ref{V_extrapolate05}, but for $N_f=2$.
}
\label{V_extrapolate2}
\end{figure}
For example, in Figs.~(\ref{V_extrapolate05})--(\ref{V_extrapolate2}) we plot the difference in the potentials $(V_{Z_N}-V_{Z_2})/N^2$ for $a_tm=0$, $a_t/a_s=0.01$ and $N_f=0.5,1,2$, versus $(2\pi/L)^3$. While $V_{Z_N}$ and $V_{Z_2}$ are expected to depend linearly on $(2\pi/L)^{4}$ and $(2\pi/L)^3$, respectively, then the fact that the plots are linear means that most of the variation in the difference $V_{Z_N}-V_{Z_2}$ reflects the variation of $V_{Z_2}$. Our data was generated for $L=70,90,100,140,160$, and as is clear from the figures, at $L\to\infty$ we see that $V_{Z_N} < V_{Z_2}$ and so that the $Z_N$ symmetry is intact. Fortunately, we see that the linear behavior sets in already at the small values of $L$ where we can perform the numerical integration at a reasonable computational cost. 
Therefore, from here on we shall restrict ourself to performing the numerical integrations with $L=60$ and $90$, and map the phase diagram according to the $L\to\infty$ linear extrapolations of these potentials. Let us emphasize, however, that while it is important to perform the large-$L$ extrapolations at small values of $a_t/a_s$, then at moderate values of this parameter the results we obtain prior to the extrapolations are quite close to their large-$L$ limit. This is expected since at very small values of $a_t/a_s$, only the vicinity of the Brillouin Zone origin is important, and a finer grid is necessary. In practice, we find that it is only at $a_s/a_t\stackrel{<}{_\sim}0.2$ that the linear extrapolations are important, while for larger $a_s/a_t$, a numerical integration with $L=90$ is already reflecting the situation at $L=\infty$.

\begin{figure}[hbt]
\centerline{
\includegraphics[width=10cm]{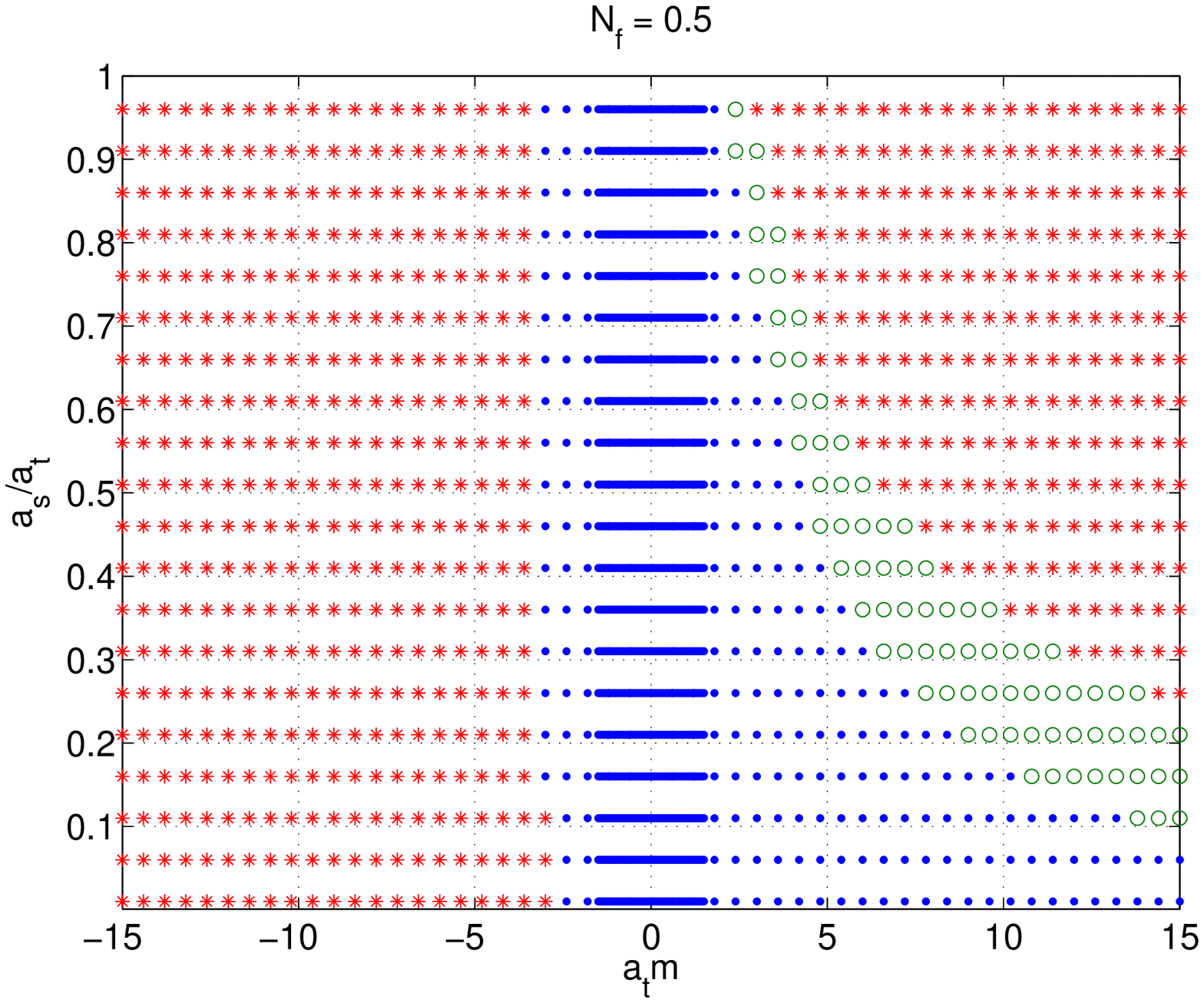}
}
\caption{A map of the phase space according to the one loop potential for anisotropic lattice spacings and $N_f=1/2$ (a single Majorana fermion in the continuum of the $Z_N$ invariant case) as a function of $a_tm$ and $a_s/a_t$. Bursts (red) denote the points in the parameter space where the $Z_N$ symmetry was seen to be completely broken, circles (green) denote the regime where the $Z_N$ symmetry is broken down to $Z_2$, and dots (blue) the regime where the $Z_N$ symmetry is intact. Note that the scan in the proximity of $a_tm=0$ was done with a finer resolution in $a_tm$. }
\label{asym05_map}
\end{figure}

\begin{figure}[hbt]
\centerline{
\includegraphics[width=10cm]{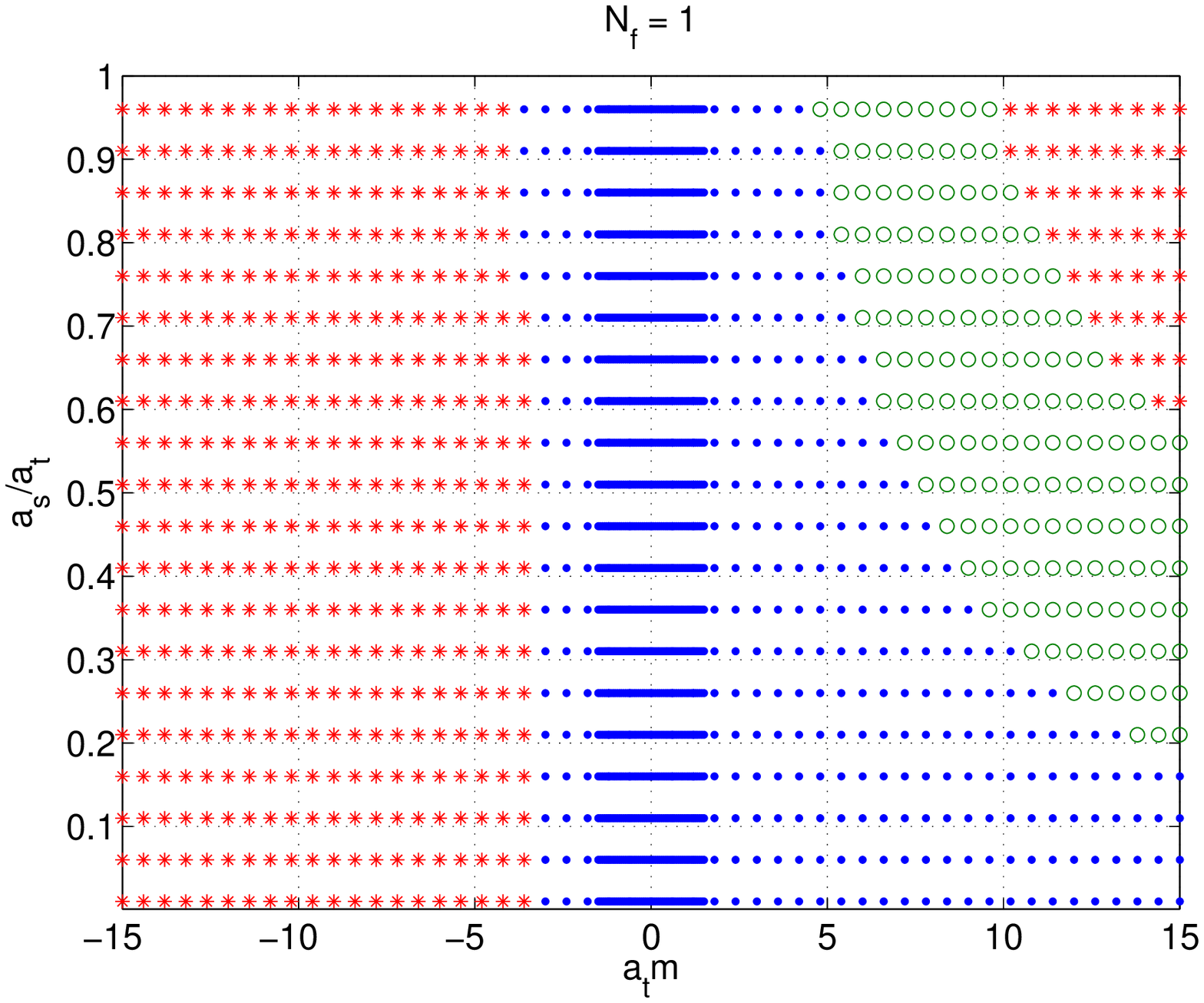}
}
\caption{As in Fig.~\ref{asym05_map} but for $N_f=1$.
}
\label{asym1_map}
\end{figure}

\begin{figure}[hbt]
\centerline{
\includegraphics[width=10cm]{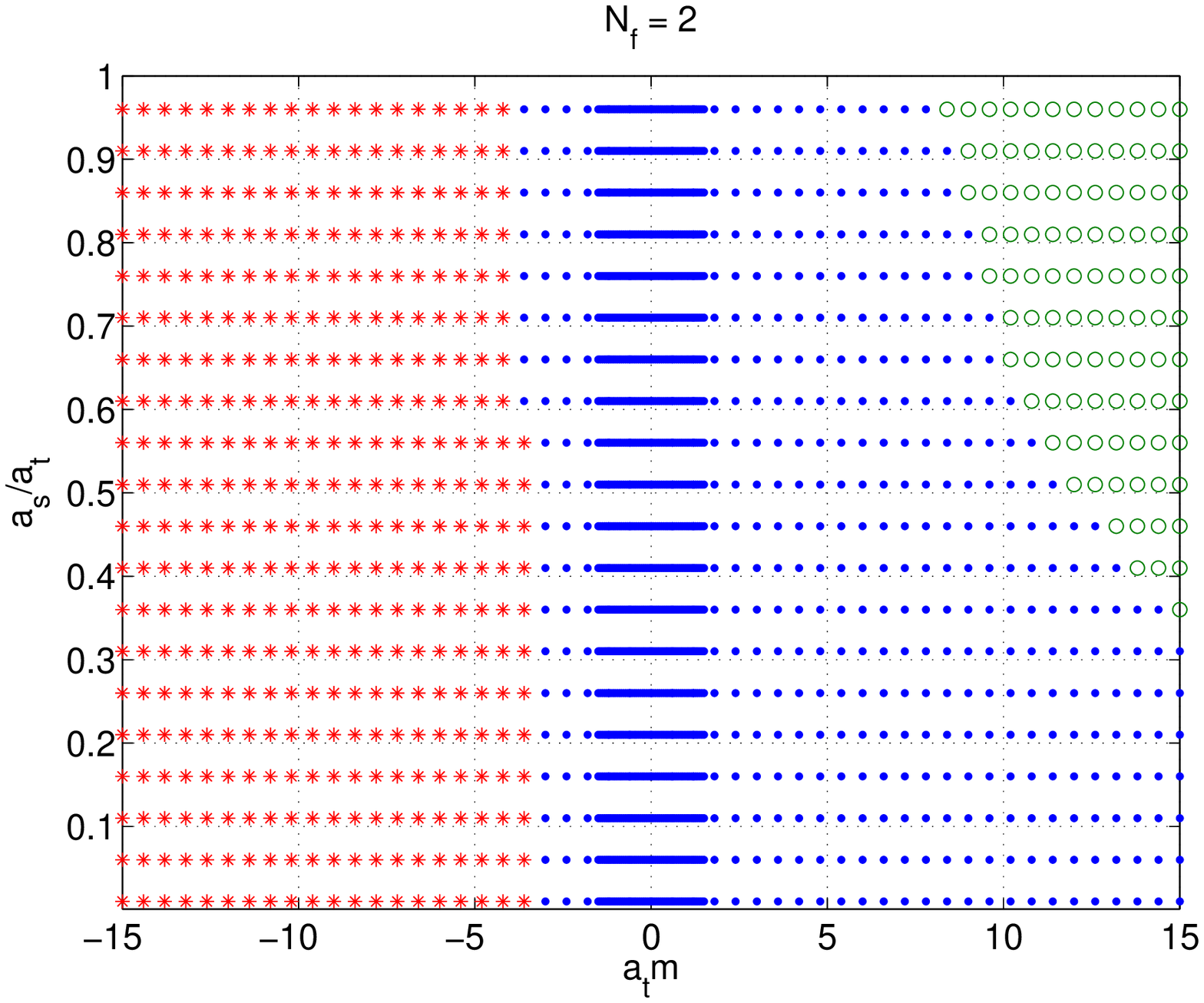}
}
\caption{As in Fig.~\ref{asym05_map} but for $N_f=2$.
}
\label{asym2_map}
\end{figure}

The results we find are quite interesting: we see that introducing an anisotropy makes the range in which the $Z_N$ symmetry is intact more extended. For physically relevant values of the quark mass, however, nothing dramatic happens and the ground state is still $Z_N$ invariant. We present the map of the phase diagrams in Figs.~(\ref{asym05_map})-(\ref{asym2_map}).

\section{Conclusions}
\label{summary}

In this paper we analyzed the large-$N$ volume independence of four dimensional QCD with adjoint fermions, and we find that it works in weak coupling if one regularizes the theory on the lattice with Wilson fermions. Specifically, we studied this regularization for different number of flavors, different quark masses, and a varying anisotropy between the spatial and temporal lattice spacings.  Our calculation is performed at one-loop and we calculate the corresponding effective potential $V$ as a function of the eigenvalues $\{e^{i\theta_a}\}$ of the holonomy in the reduced direction (the Polyakov loop $P$). We find that $V(\theta)$ prefers a $Z_N$ invariant ground state for moderately light (and even quite heavy) fermions, and for most values of the anisotropy. Our results hold for $1/2,1,2$ Dirac flavors (one, two and four Majorana fermions in the continuum). If, however, the fermions are extremely heavy, then the reduced theory spontaneously breaks its $Z_N$ symmetry and large-$N$ volume independence breaks down. We also see that the $Z_N$ symmetry is broken in the (physically uninteresting) super-critical regime of the lattice Wilson theory. 

Another goal of this paper was to understand whether there is any tension between the results of Ref.~\cite{BBCS}, which treated the volume-reduced model as a $3D$ continuum effective field theory (EFT), and those of Ref.~\cite{AEK} (that worked directly with the $4D$ gauge theory in the continuum), and whether the results of the former signal any problems with large-$N$ reduction. In particular, for massless quarks Ref.~\cite{AEK} finds that $V(\theta)$ has a $Z_N$ invariant vacuum, while Ref.~\cite{BBCS} sees a vacuum that breaks the $Z_N$ symmetry down to $Z_2$. In fact, from the previous paragraph it seems that the results in Ref.~\cite{BBCS} also contradict the results of our lattice calculation, in which the $3D$ spatial dimensions of the reduced model are discretized to have a finite lattice spacing $a_s$.


Before we describe how we resolved these `contradictions', we wish to emphasize that treating the reduced model as a three-dimensional EFT may be useful but is not necessary.\footnote{The reason why Ref.~\cite{BBCS} choose to do so is because it may open a window for using three dimensional analytic techniques to study the four dimensional theory.} In particular, in the lattice calculation performed in this paper we treat the $L_t=1$ reduced model as a theory that is defined with a fixed cutoff, or equivalently that has finite bare lattice parameters. For values of lattice parameters where the center symmetry is intact, the theory is large-$N$ equivalent to a corresponding theory with the same bare lattice parameters, the same field content, and the same cutoff, but with $L_t=\infty$. This equivalence is true not only at low energies, but all the way up to the cutoff scale. Removing the cutoff $a_s$ from this construction is finally done  {\em after taking the large-$N$ limit}, and according to the RG equations of the $4D$ large-$N$ theory. 

While our approach is the standard way one defines large-$N$ reduction nonperturbatively \cite{EK,GK}, it does not mean that the approach of Ref.~\cite{BBCS} is not useful, and we still need to understand how to resolve the apparent contradiction between the results of this approach and what we and Ref.~\cite{AEK} find for the symmetry of the ground state. For that purpose we first showed in Section~\ref{EFT} that the three-dimensional EFT defined by the reduced model is non-renormalizable. One of the consequences of this is a set of linear divergences in $V(\theta)$ that depend on $\theta$. These are radiatively generated as mass terms for the Polyakov loops and from the point of view of EFT we need to cancel them.  Thus, one needs to add to the action of the EFT certain counter terms. These turn out to be relevant operators and after canceling the UV divergences, they leave us with extra finite additions to the one loop potential that depend on new low energy constants (LEC). Specifically,  we showed that the finite contribution of the counter terms is of the form
\begin{equation}
\delta S_{\rm finite} \sim  \int d^3x \left\{b_1\, |\tr P(x)|^2+b_2|\tr P^2(x)|^2\right\} .\label{missing1}
\end{equation}
where the LEC $b_1$ and $b_2$ have mass dimension three. We also identified the  counter terms for general values of $L_t$ and of the spatial dimension $D$ (see Section~\ref{EFT}). 
Interestingly, the counter terms constitute a subset of the operators that Ref.~\cite{DEK} suggested to add to the Eguchi-Kawai model in order to make its ground state $Z_N$-invariant. In this paper we see that in the EFT approach, their presence is dictated by the regularization process of the reduced model. (Note that if the theory has a ground state with a $Z_N$ symmetry, then these operators, when projected back to $4D$, do not affect the leading large-$N$ dynamics -- again see Ref.~\cite{DEK}).      

Therefore, the reduced model, seen as a three-dimensional EFT, is defined not only by parameters like the quark mass $m$ and the number of flavors $N_f$, but also by $b_1$ and $b_2$. This tells us what is the cause for the different results one obtains in different regularizations: they reflect different choices for $b_1$ and $b_2$. To make this clear let us detail the choices made for these LEC by the different calculations.
\begin{enumerate}[I.]
\item \underline{The three dimensional EFT continuum calculation of Ref.~\cite{BBCS}:}

Here the subtraction of the infinities was done with minimally subtracted dimensional regularization (MSDR). This choice effectively sets power law divergences to zero and replaces them by finite $\theta$-dependent functions. The fact that $\delta S_{\rm finite}$ from \Eq{missing1} was not added to the action studied in Ref.~\cite{BBCS} means that MSDR actually fixed $b_1=b_2=0$. This choice is made implicitly by the regulator and is why MSDR is considered `dangerous' when dealing with power-law divergences \cite{Luscher}.
\item \underline{The lattice calculation presented in our work (Section~\ref{recap}):}

Above we emphasized we do not treat the reduced model as an EFT. Despite this, and to understand the difference between our results and those of Ref.~\cite{BBCS}, we momentarily choose to depart this approach, and think about our lattice as a regulator for a $3D$ EFT which is alternative to MSDR. It is easy to see that, from this point of view, the lattice fixed $b_{1}$ and $b_{2}$ to specific values that scale like $1/a_s$ for small $a_s$ (because we did not add \Eq{missing1} to our lattice action -- for details see Section~\ref{meaningBBCS}).
\end{enumerate}

There is absolutely no reason to expect that the implicit choices made for $b_{1,2}$ in MSDR and on the lattice should lead to the same physical results.  In particular, it is quite possible that these choices lead to different realizations of the $Z_N$ symmetry in the ground state. Indeed this is what happens: the choice made by MSDR tends to break  the $Z_N$ symmetry, while the one of the lattice (with Wilson fermions) tends to preserve it.

To obtain the same physical results in different regularizations, one would need to fix the physical parameters ($m$, $N_f$, etc.), explicitly introduce the terms in \Eq{missing1} into $V(\theta)$, and tune the values of $b_1$ and $b_2$ in a regularization dependent way (instead of letting the regulator fix them implicitly). Then, there will be some choice within MSDR,  $b_{1,2}=b^{\rm MSDR }_{1,2}$, that will give the same physical results that a different choice, $b_{1,2}=b^{\rm lattice}_{1,2}$, gives on the lattice.

 Moreover, to make a regularization-independent statement on the {\em absence} of $Z_N$ symmetry in the ground state of the theory, one needs to show that, within a certain regularization, the center symmetry breaks for all values of $b_1$ and $b_2$.
As we say above, this was not included in the analysis of Ref.~\cite{BBCS} and so the result of that work does {\em not} mean that large-$N$ reduction of QCD with light adjoint fermions is invalid. 
Put differently,  Ref.~\cite{BBCS} effectively studied $V(\theta)$ in a subspace of the full parameter space of the EFT; a subspace that turns out to have a broken $Z_N$ symmetry for physically relevant values of the quark mass. If we view our lattice calculation as an EFT, then the choices made for $b_{1,2}$ by the lattice can also be seen as a restriction to a subspace in parameter space. But in contrast to what happens with MSDR, in our case the subspace defined by $b^{\rm lattice}_{1,2}$ turned out to generically have a $Z_N$-invariant ground state for $V(\theta)$. 

Is it possible to change the results of Ref.~\cite{BBCS} by making its choice of $b_{1,2}$ nonzero? The answer seems to be yes. For example, for $m=0$ and $N_f=1$, Ref.~\cite{BBCS} reported a breakdown of $Z_N\to Z_2$, and it is fairly clear that by increasing both $b_1$ and $b_2$ to sufficiently positive values, one can get rid of this symmetry breakdown. Indeed, this fact is what makes the results of Ref.~\cite{BBCS} consistent with the ones we present in the current paper, namely that large-$N$ reduction works for a physically relevant range of the parameters in the gauge theory. This also means that there is no tension between the results of Ref.~\cite{BBCS} and those of Ref.~\cite{AEK}. 

\bigskip

The fact that we can view the lattice definition of the $L_t=1$ reduced model as a regularization of a three-dimensional nonrenormalizble EFT with fixed $b_{1,2}$ teaches us the following important lesson. Other lattice constructions of the $L_t=1$ model (for example ones similar to \cite{CD} which use staggered fermions, or any other type of fermions) can be seen as alternative regulators for the same EFT. Then, if we do not add the terms in \Eq{missing1} to their action, they will implicitly choose their own values for $b_{1,2}$. Importantly, it is not guaranteed that these choices will generically lead to a $Z_N$ symmetric ground state in the physically interesting regime. In that sense, the result we present in this paper for Wilson fermions cannot be anticipated in advance. Clearly, this means that it will be wise to perform one-loop calculations of the form we did in this paper for each regularization of a single-site model, prior to its (computationally costly) numerical Monte-Carlo study.

\bigskip

The question of whether the $Z_N$ symmetry is intact in our regularization also at moderate couplings, where the one-loop calculation is unreliable, can be answered only via non-perturbative Monte-Carlo simulations. For example, in the $N_f=2$ case, the study in Ref.~\cite{CD} suggests that the answer is rather complicated and may be sensitive to the bare lattice parameters. The results we present in Section~\ref{asym} on the phase structure in the $a_s/a_t-a_tm$ plane can be viewed as another example of a moderate case of this sensitivity, which is harmless for the large-$N$ reduction program.

\bigskip

Indeed, in the companion publication \cite{BS}  non-perturbative Monte-Carlo simulations were used to explore large-$N$ reduction and find evidence that for some values of the lattice coupling and quark masses, the theory with symmetric lattice spacings can be successfully reduced to a single site in {\em all} the euclidean directions.
Anticipating this result using a one loop calculation is, unfortunately, not straightforward, and compared to the one-loop calculation presented in this paper, is complicated by  IR divergences. These need to be taken into account in a similar fashion to the way Ref.~\cite{singular} estimated the free energy of singular tolerons. Nonetheless, the values of the quark mass at which we see a transition from a $Z_N$ broken phase into a $Z_N$ invariant phase in the Monte-Carlo simulations of Ref.~\cite{BS} are in qualitative agreement with the values we find analytically in the current paper.


\begin{acknowledgements}
For many useful discussions, correspondences, and comments on the manuscript, I thank Carlos Hoyos, Paulo Bedaque, Herbert  Neuberger (who also gave us a copy of his unpublished paper \cite{HN}), Steve Paik, Steve Sharpe, Mithat \"Unsal, and Larry Yaffe. This study was supported in part by the U.S. Department of Energy under Grant No. DE-FG02-96ER40956.
\end{acknowledgements}



\begin{thebibliography}{99}

\bibitem{EK}
  T.~Eguchi and H.~Kawai,
  Phys.\ Rev.\ Lett.\  {\bf 48}, 1063 (1982).

\bibitem{AEK}
  P.~Kovtun, M.~Unsal and L.~G.~Yaffe,
  JHEP {\bf 0706}, 019 (2007)
  [arXiv:hep-th/0702021].

\bibitem{Yaffe_coherent}
  L.~G.~Yaffe,
  Rev.\ Mod.\ Phys.\  {\bf 54}, 407 (1982).

\bibitem{BHN}
  G.~Bhanot, U.~M.~Heller and H.~Neuberger,
  Phys.\ Lett.\  B {\bf 113}, 47 (1982).

\bibitem{MK}
  V.~A.~Kazakov and A.~A.~Migdal,
  Phys.\ Lett.\  B {\bf 116}, 423 (1982).

\bibitem{KNN}
  J.~Kiskis, R.~Narayanan and H.~Neuberger,
  Phys.\ Lett.\  B {\bf 574}, 65 (2003)
  [arXiv:hep-lat/0308033].

\bibitem{HN_lat05}
  R.~Narayanan and H.~Neuberger,
  PoS {\bf LAT2005}, 005 (2006)
  [arXiv:hep-lat/0509014].

\bibitem{DEK}
  M.~Unsal and L.~G.~Yaffe,
  Phys.\ Rev.\  D {\bf 78}, 065035 (2008)
  [arXiv:0803.0344 [hep-th]].


\bibitem{QEK}
  B.~Bringoltz and S.~R.~Sharpe,
  Phys.\ Rev.\  D {\bf 78}, 034507 (2008)
  [arXiv:0805.2146 [hep-lat]].
  B.~Bringoltz and S.~R.~Sharpe,
  PoS {\bf LATTICE2008}, 055 (2008)
  [arXiv:0810.1239 [hep-lat]].


\bibitem{ASV}
  A.~Armoni, M.~Shifman and G.~Veneziano,
  Nucl.\ Phys.\  B {\bf 667}, 170 (2003)
  [arXiv:hep-th/0302163].


\bibitem{BBCS}
  P.~F.~Bedaque, M.~I.~Buchoff, A.~Cherman and R.~P.~Springer,
  arXiv:0904.0277 [hep-th].

\bibitem{BS} B.~Bringoltz and S.~R.~Sharpe,
in preparation.



\bibitem{HN} H.~Neuberger,
Unpublished, July 2001.


\bibitem{asym_papers}
  R.~G.~Edwards, B.~Joo and H.~W.~Lin,
  Phys.\ Rev.\  D {\bf 78}, 054501 (2008)
  [arXiv:0803.3960 [hep-lat]].


\bibitem{Rothe}
  H.~J.~Rothe,
  World Sci.\ Lect.\ Notes Phys.\  {\bf 74} (2005) 1.

\bibitem{GKPZ}
  D.~J.~Gross and Y.~Kitazawa,
  Nucl.\ Phys.\  B {\bf 206}, 440 (1982).
  G.~Parisi and Y.~C.~Zhang,
  Phys.\ Lett.\  B {\bf 114}, 319 (1982).

\bibitem{0modes}
  H.~Aoki, S.~Iso, H.~Kawai, Y.~Kitazawa and T.~Tada,
  Prog.\ Theor.\ Phys.\  {\bf 99} (1998) 713
  [arXiv:hep-th/9802085].
\bibitem{ABL}
  T.~Appelquist and C.~W.~Bernard,
  Phys.\ Rev.\  D {\bf 22}, 200 (1980).
  T.~Appelquist and C.~W.~Bernard,
  Phys.\ Rev.\  D {\bf 23}, 425 (1981).
  A.~C.~Longhitano,
  Phys.\ Rev.\  D {\bf 22}, 1166 (1980).



\bibitem{BUP}
  T.~Banks and A.~Ukawa,
  Nucl.\ Phys.\  B {\bf 225}, 145 (1983).
  R.~D.~Pisarski,
  Phys.\ Rev.\  D {\bf 74}, 121703 (2006)
  [arXiv:hep-ph/0608242].



\bibitem{AHCG}
  N.~Arkani-Hamed, A.~G.~Cohen and H.~Georgi,
  Phys.\ Rev.\ Lett.\  {\bf 86}, 4757 (2001)
  [arXiv:hep-th/0104005].
  N.~Arkani-Hamed, A.~G.~Cohen and H.~Georgi,
  Phys.\ Lett.\  B {\bf 513}, 232 (2001)
  [arXiv:hep-ph/0105239].

\bibitem{GL}
  J.~Gasser and H.~Leutwyler,
  Annals Phys.\  {\bf 158}, 142 (1984).


\bibitem{PDS}
  D.~B.~Kaplan, M.~J.~Savage and M.~B.~Wise,
  Phys.\ Lett.\  B {\bf 424}, 390 (1998)
  [arXiv:nucl-th/9801034].

\bibitem{Luscher}
  M.~Luscher,
  Nucl.\ Phys.\  B {\bf 219}, 233 (1983).

\bibitem{Aoki}
  M.~Golterman, S.~R.~Sharpe and R.~L.~.~Singleton,
  Nucl.\ Phys.\ Proc.\ Suppl.\  {\bf 140}, 335 (2005)
  [arXiv:hep-lat/0409053].


\bibitem{CD}
  G.~Cossu and M.~D'Elia,
  arXiv:0904.1353 [hep-lat].

\bibitem{GK}
  D.~J.~Gross and Y.~Kitazawa,
  Nucl.\ Phys.\  B {\bf 206}, 440 (1982).

\bibitem{singular}
  A.~Coste, A.~Gonzalez-Arroyo, J.~Jurkiewicz and C.~P.~Korthals Altes,
  Nucl.\ Phys.\  B {\bf 262} (1985) 67.




\end{thebibliography}
\end{document}